\documentclass[useAMS,usenatbib]{mn2e}

\def\prl{Phys. Rev. Lett.}

\def\apjl{ApJL}
\def\na{NewA}

\def\mnras{MNRAS}
\def\apj{ApJ}

\def\prd{PhysRevD}
\def\jcap{JCAP}
\def\physrep{Phys. Rept.}
\def\araa{Annual Review of Astronomy and Astrophysics}

\def\dgemail{gilmanda@ucla.edu}

\usepackage[titletoc]{appendix}
\usepackage{graphicx}
\usepackage{float}
\usepackage{amssymb}
\usepackage{amsfonts}
\usepackage{amsmath} 
\usepackage{color}
\usepackage{wrapfig}
\usepackage{hyperref}
\usepackage{algorithm}
\usepackage{algpseudocode}
\usepackage{physics}

\usepackage{natbib}

\def\dlos{{\delta_{\rm{los}}}}

\def\msun{{\rm{M}_{\odot}}}

\def\a0{{a_0}}
\def\mhm{{m_{\rm{hm}}}}

\title[Flux ratio statistics with line of sight halos]{Probing dark matter structure down to $10^7$ solar masses: flux ratio statistics in gravitational lenses with line of sight halos} 
\author[Gilman et al.]{\parbox{\textwidth}{
		Daniel Gilman$^{1}$ \thanks{\dgemail}, Simon Birrer $^{1}$ , Tommaso Treu $^{1}$, Anna Nierenberg $^{2}$, Andrew Benson $^{3}$}
	\\
	\\
	\parbox{\textwidth}{
		$^{1}$Department of Physics and Astronomy, University of California,
		Los Angeles, CA 90095, USA\\
		$^{2}$Jet Propulsion Laboratory, California Institute of Technology, 4800 Oak Grove Dr, Pasadena, CA 91109, USA\\
		$^{3}$Carnegie Observatories, 813 Santa Barbara Street, Pasadena, CA 91101, USA
	}
}

\begin{document}
	
	\voffset-.6in
	
	\date{Accepted . Received }
	
	\pagerange{\pageref{firstpage}--\pageref{lastpage}} 
	
	\maketitle	
	\label{firstpage}
	\begin{abstract}
		Strong lensing provides a powerful means of investigating the nature of dark matter as it probes dark matter structure on sub-galactic scales. We present an extension of a forward modeling framework that uses flux ratios from quadruply imaged quasars (quads) to measure the shape and amplitude of the halo mass function, including line of sight (LOS) halos and main deflector subhalos. We apply this machinery to 50 mock lenses --- roughly the number of known quads --- with warm dark matter (WDM) mass functions exhibiting free-streaming cutoffs parameterized by the half-mode mass $m_{\rm{hm}}$. Assuming cold dark matter (CDM), we forecast bounds on $m_{\rm{hm}}$ and the corresponding thermal relic particle masses over a range of tidal destruction severity, assuming a particular WDM mass function and mass-concentration relation. With significant tidal destruction, at $2 \sigma$ we constrain $m_{\rm{hm}}<10^{7.9} \left(10^{8.4}\right) M_{\odot}$, or a 4.4 (3.1) keV thermal relic, with image flux uncertainties from measurements and lens modeling of $2\% \left(6\%\right)$. With less severe tidal destruction we constrain $m_{\rm{hm}}<10^{7} \left(10^{7.4}\right) M_{\odot}$, or an 8.2 (6.2) keV thermal relic. If dark matter is warm, with $m_{\rm{hm}} = 10^{7.7} M_{\odot}$ (5.1 keV), we would favor WDM with $m_{\rm{hm}} > 10^{7.7} M_{\odot}$ over CDM with relative likelihoods of 22:1 and 8:1 with flux uncertainties of $2\%$ and $6\%$, respectively. These bounds improve over those obtained by modeling only main deflector subhalos because LOS objects produce additional flux perturbations, especially for high redshift systems. These results indicate that $\sim 50$ quads can conclusively differentiate between warm and cold dark matter.  
	\end{abstract}
	
	\section{Introduction}
	Theories of particle dark matter predict that the enigmatic particle(s) collect in gravitationally bound halos. The mass function and density profiles of these objects depend on the particle nature of dark matter itself. For example, theories with cold dark matter predict an abundance of low mass halos, and cuspy $r^{-1}$ central density profiles \citep{Moore++99,Springel++08,Fiacconi++16}. In warm dark matter (WDM) models, diffusion of dark matter particles in the early universe wipes out density fluctuations below a characteristic scale that depends on the production mechanism of the WDM particle candidate \citep{Kusenko09,ShoemakerKus09,Abazajian17}. Suppression of small-scale power in WDM models results in a turnover in the halo mass function and a dearth of small-scale structure at later times \citep{Bode++01,Schneider++12,Lovell++14}. In self-interacting dark matter (SIDM) theories, scattering between dark matter particles produces cored density profiles in individual halos \citep{S+S00,Rocha++13,Vogelsberger++16,Kamada++17,TulinYu18}. Finally, in `fuzzy' dark matter scenarios the kpc-scale de Broglie wavelength of ultra-light dark matter particles results in quantum mechanical phenomena on galactic scales, which produces large soliton cores \citep{Hui++17,Robles++19}. To date, the strongest constraints on WDM come from the Lyman-$\alpha$ forest \citep{Viel13, Irsic++17}, while cosmological probes on large scales \citep{CyrRacine++14,Bringmann++17} and in galaxy clusters \citep{Kim++17,Andrade++19} constrain the interaction cross section in SIDM models.
	
	Two challenges to CDM in particular spur interest in alternative theories. First, natural CDM particle candidates have not yet been detected, despite decades of experimental searches \citep{Aprile++18}. Second, the suppression of small scale structure in WDM, and cored density profiles associated with SIDM, possibly alleviate tension between observations and the predictions on sub-galactic scales, dubbed the `Small-Scale Crisis' of CDM \citep[see][and references therein]{BullockBK17}. 
	
	Traditional astrophysical challenges to the CDM model, however, are predicated on assumptions related to baryonic physics. This has the undesirable consequence of propagating uncertainties from sub-galactic astrophysics onto inferences of dark matter properties, and results in covariance between baryonic astrophysics and dark matter physics. Supernova and stellar feedback inside halos, for instance, and the tidal destruction of subhalos by their host galaxy, mimic the observable signatures of SIDM and WDM models, respectively \citep{Tollet++16,Read++18, DespVeg16,GK++17,Kim++18,Despali++18b}. Moreover, in some cases, the uncertainties related to baryonic astrophysical processes can be larger than the differences between CDM, WDM, and SIDM \citep[e.g.][]{Nie++16}. To isolate dark matter physics from sub-galactic astrophysics, and to differentiate between CDM, WDM, and SIDM, one must look to masses below $10^8 \msun$, where subhalos are expected to be devoid of stars and completely dark in the case of CDM, or absent in the case of WDM. 
	
	Gravitational lensing offers a direct probe of this elusive, low-mass regime. It circumvents the complications associated with using luminous matter to trace the dark matter by enabling the direct measurement of the distribution of matter across cosmological distance, and is sensitive to mass scales where astrophysical effects are thought to be too weak to significantly alter the structure of halos. It also compliments other probes of dark matter, such as the Lyman-$\alpha$ forrest, since lensing depends on different systematics and measures the halo mass function directly. 
	
	Ultimately, analysis of strong lenses hinges on separating mass distributions that vary on large scales (the lensing galaxy and its parent dark matter halo) from small scale structure in the main lens plane and along the line of sight. In strong lens systems with luminous arcs, the analysis consists of iteratively fitting a smooth model to the flux in pixels of an image while simultaneously reconstructing the background source. This process can reveal the presence of small scale structure in the arcs \citep[see e.g.][]{Veg++14,Hezaveh++16,Vegetti++18,Ritondale++18}. \citet{Birrer++17a} performed this analysis, and placed constraints on the free streaming length of dark matter. Recently, several authors have proposed using the surface brightness residuals from lens models fit to luminous arcs and to infer the power spectrum of dark matter in strong lenses \citep{Hezaveh++16b,DiazRivero++18,Cyr-Racine++18}, and \citet{Bayer++18} applied this method to a strong lens system.  
	
	In addition to extended arcs, some strong gravitational lenses produce four images (quads) of an unresolved background source, such as a quasar. The magnification ratios (flux ratios) between multiple images of unresolved sources have long been recognized as powerful probes of small scale structure near lensed images \citep{MaoSchnieder98,MetcalfMadau01}, and have been used to test the predictions of CDM and to detect structure near individual objects \citep{D+K02,Amara++06,Xu++12,FadleyKeeton12,MacLeod++13,Nierenberg++14,Xu++15,Nierenberg++17}. Recently, \citet{Nierenberg++14,Nierenberg++17} used image flux ratios measured from narrow line emission, a method first proposed by \citep{MoustakasMetcalf02}, to study substructure in strong lenses. The significance of this advance derives from the fact that the magnification of a lensed image is a function of background source diameter \citep{DoblerKeeton02}; the narrow-line region, which typically subtends angles on scales of a few tens of milliarcseconds, is resilient to contaminating effects of microlensing by stars, while still being sensitive to the milliarcsecond perturbations sourced by dark matter halos above $10^6 \msun$ with current astrometric precision of a few m.a.s. \citep{Nierenberg++17}. 
	
	In this work, we extend the formalism presented by \citep{Gilman++18} to include the contribution from dark matter halos along the line of sight. Since field halos do not orbit in a steep galactic potential with star formation, stellar feedback, and other complications, they constitute an ideal laboratory for studying the intrinsic structure of dark matter halos. Several studies investigate the role of the line of sight halos on flux ratio perturbations in strong lenses  \citep{Chen++03,Metcalf05,MirandaMaccio07,Xu++12,Inoue++12}, and \citet{Despali++18} address the line of sight contribution in the context of gravitational imaging with luminous arcs. The consensus from these works is that the line of sight halos affect lensing observables, possibly becoming the dominant source of perturbation to smooth lens models for lenses at high redshift.  
	
	The analysis presented here builds on previous analysis of multiple image lenses in several ways. First, we quantify the signal from non-linear multi-plane lensing effects on flux ratios with finite-size background sources, and combine this multi-plane lensing machinery with a forward-generative model to measure the shape and amplitude of the halo mass function by combining flux ratio statistics from a sample of lenses. We also marginalize over parameters such as the background source size and the power law profile of the main deflector, both of which can affect the flux ratios between images. We demonstrate how well this method constrains the free-streaming length of dark matter in the presence of uncertainties associated with measurements and lens modeling, and apply the machinery to a set of 50 simulated quads. The number 50 is chosen since it is roughly the size of the current sample of known quads \citep[][HST GO-15652]{Shajib++19} with a similar distribution of lens and source redshifts. 
	
	This paper is organized as follows: First in Section \ref{sec:rendering}, we describe our prescription for modeling the line of sight halo mass function, and the subhalo mass function in the main lens plane. In Section \ref{sec:multiplanelensing}, we discuss the impact of line of sight halos on flux ratio observables. In Section \ref{sec:simsetup}, we describe the forward modeling procedure implemented in the simulations, and the process for creating mock datasets. Finally, in Section \ref{sec:results} we present the results of simulations run with a mock data set in which we infer dark matter and lens model parameters with a Bayesian framework. Finally, Section \ref{sec:conclusion} summarizes our main results. All lensing computations performed in this work utilize the open-source gravitational lensing software {\textsc{lenstronomy}} \citep{Birrer++15,BirrerAmara18}. Cosmological calculations, in particular the line of sight halo mass function and two-halo term, are computed with the software package {\textsc{colossus}} \citep{Diemer17}. We assume a flat cosmology with parameters from WMAP9 \citep{WMAP9cosmo}: $\sigma_8 = 0.82$, matter density $\Omega_{m} = 0.28$ and hubble constant $h = 0.7$. When quoting halo masses, we refer to $M_{200}$ computed with respect to the critical density of the universe at z=0.
	
	\section{Modeling the  line of sight and subhalo mass functions}
	\label{sec:rendering}
	
	This section describes the parameterization of the subhalo mass function in the main deflector, and the halo mass function along the line of sight, as well as the density profile for individual halos. We then describe our parameterization of free-streaming effects in WDM models, both on the mass functions and the mass-concentration relation. The forward model, described in Section \ref{sec:simsetup}, will use these parameterizations to render realistic populations of dark matter structure for lensing computations.
	
	\subsection{Mass profile of individual halos}
	\label{ssec:mdefs}
	\begin{figure}
		\includegraphics[clip,trim=0cm 0cm 0cm
		0cm,width=.48\textwidth,keepaspectratio]{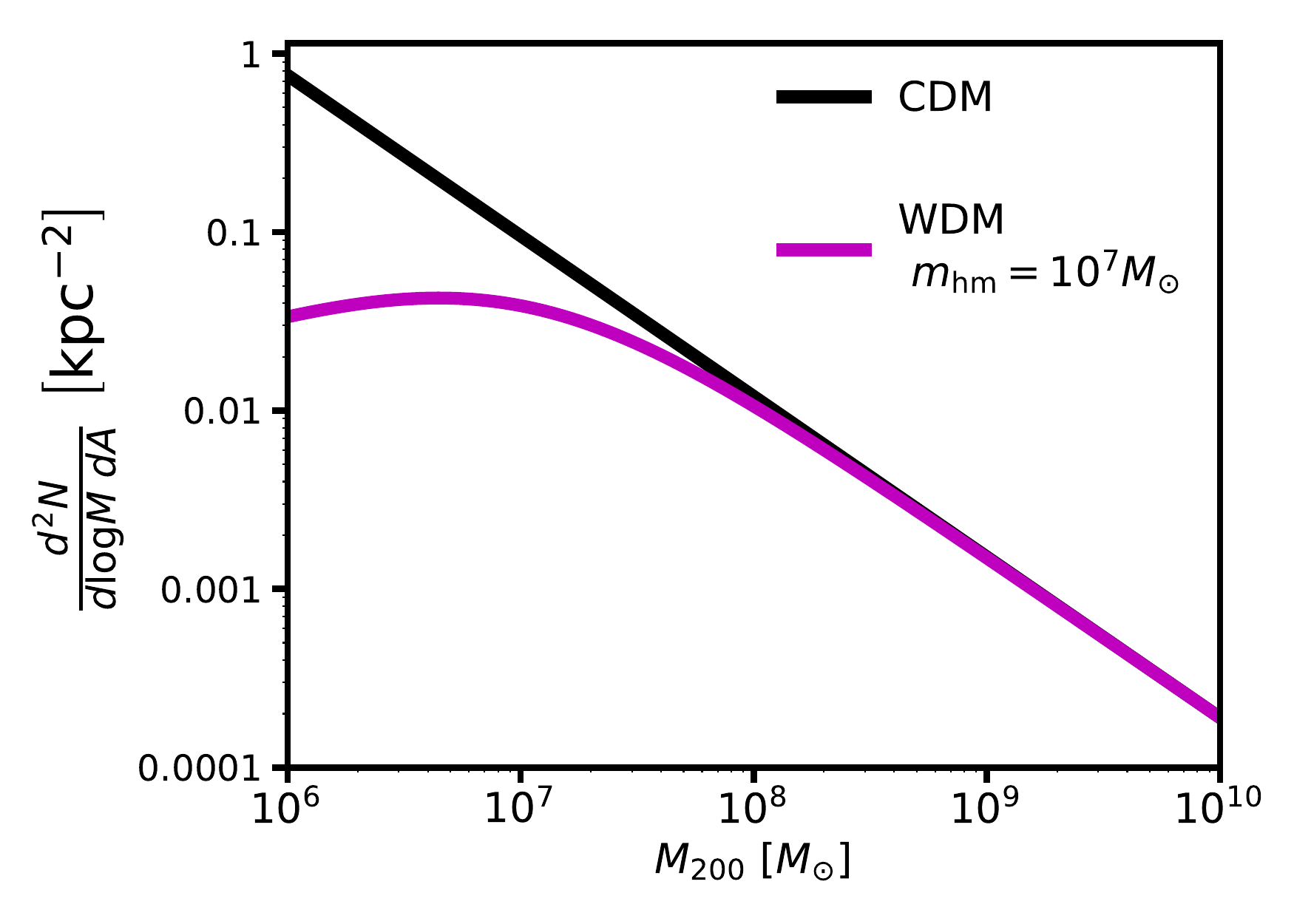}
		\includegraphics[clip,trim=0cm 0cm 0cm
		0cm,width=.48\textwidth,keepaspectratio]{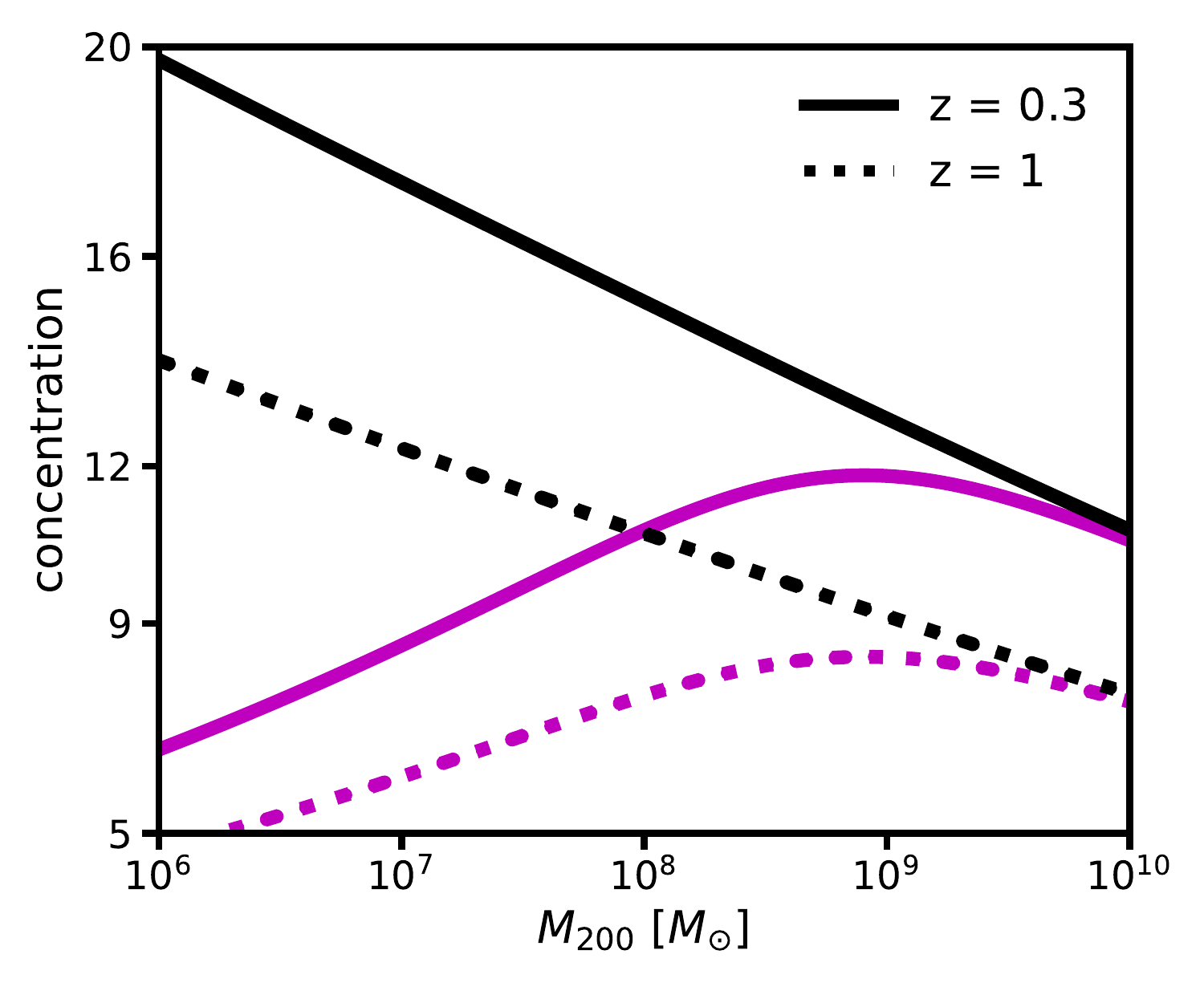}
		\caption{\label{fig:mcrelation} {\bf{Top:}} The subhalo mass function for CDM and a WDM scenario with a half-mode mass of $10^7 \msun$. The line of sight halo mass function looks qualitatively similar, but evolves slightly with redshift. {\bf{Bottom:}} The mass concentration relation for the same CDM and WDM models as in the upper panel. The effects of free-streaming on the mass-concentration relation alter the properties of halos two orders of magnitude above the half-mode mass.}
	\end{figure}	
	We model the density profiles of dark matter halos using truncated NFW profiles \citep{Baltz++09} 
	\begin{equation}
		\label{eqn:massprofile}
		\rho \left(r, r_s, r_t\right) = \frac{\rho_0}{x \left(1+x\right)^2} \frac{\tau^2}{x^2 + \tau^2}
	\end{equation}
	where $\tau = \frac{r_t}{r_s}$ and $x = \frac{r}{r_s}$.
	
	In the main lens plane, we tidally truncate subhalos through a Roche limit approximation, assuming a roughly isothermal mass profile for the main lens halo mass distribution. This truncation corresponds to a scaling $r_t \propto \left(M_{200} r_{\rm{3D}}^{2}\right)^{\frac{1}{3}}$  \citep{Tormen++98,Cyr-Racine++16}. We truncate according to this scaling using the expression
	
	\begin{equation} \label{eqn:truncrad2}
		r_t = 0.68 \left(\frac{M_{200}}{10^6 \msun}\right)^{\frac{1}{3}} \left(\frac{r_{\rm{3D}}}{100 \rm{kpc}}\right)^{\frac{2}{3}} \left[\rm{kpc}\right].
	\end{equation}
	This results in a skewed distribution of $\tau$ with mean $\langle \tau \rangle \sim 6$, and a tail extending to $\tau \sim 20$.
	
	We truncate line of sight halos at $r_{50}$, or the radius where the mean enclosed density is $50 \rho_{\rm{crit}}$ \footnote{We introduce this truncation to keep the total mass per unit volume along the line of sight finite, since the mass of an NFW profile diverges. Since $r_{50}$ is much larger than the scale radius of an NFW halo, this truncation negligibly impacts observables.}. Finally, we adopt the mass-concentration-redshift relation for CDM halos presented by \citep{DiemerJoyce18} with a scatter of 0.13 dex \citep{Dutton++14}. We render halos and subhalos in the range $10^6 - 10^{10} \msun$, which captures perturbations from the smallest subhalos that affect image magnifications, given the source sizes we model. We discuss the rationale for using this mass range in Section \ref{ssec:assumptionsandpriors}.
	
	\subsection{The line of sight halo mass function}
	
	We model line of sight structure using the mass function of Sheth and Tormen \citep{ST99}, plus a boost from the 2-halo term at a distance $r$ from the main deflector $\xi_{\rm{2halo}}\left(r,M,z\right)$, where $M$ denotes the halo mass of the parent dark matter halo. The two-halo term accounts for the correlated structure in the vicinity of the main lens halo. To leading order, this term rescales the background density and the amplitude of the halo mass function. The inclusion of $\xi_{\rm{2halo}}$ results in a roughly $5-15 \%$ boost in the number of halos located at approximately the main lens redshift, depending on the normalization of the subhalo mass function and the lens redshift. We review the form of the two-halo term and its implementation in lensing simulations in Appendix \ref{app:A}. 
	
	We introduce a rescaling factor $\dlos$ to account for theoretical uncertainty regarding the amplitude of the halo mass function. This term also accounts for statistical fluctuations around the mean density of the universe, which may lead to modestly over-dense or under-dense lines of sight to individual lenses. We note, however, that due to the vast cosmological distances probed by strong lenses (order Gpc, versus kpc-scale dark matter halos and filament diameters) the dark matter structure in these volumes should be well represented by the average halo mass function in the universe which corresponds to $\delta_{\rm{los}} = 1$, modulo uncertainties in parameters such as $\sigma_8$ and $\Omega_{m}$.
	
	With these modifications, the line of sight halo mass function takes the form
	\begin{equation}
		\label{eqn:losmfunc}
		\frac{d^2N_{\rm{los}}}{dm  dV} = \dlos \left(1+ \xi_{\rm{2halo}} \right) \frac{d^2N}{dm  dV} \big \vert_{\rm{ShethTormen}}.
	\end{equation}
	This mass function yields accurate counts of isolated halos over a wide mass range. We do not model the subhalos of these objects along the line of sight, subsuming the possible effects of these small perturbers into the marginalization over $\delta_{\rm{los}}$. Line of sight halos are distributed in a double-cone geometry with opening angle $3 R_{\rm{Ein}}$, where $R_{\rm{Ein}}$ is the Einstein radius of a given lens, and a closing angle behind the main lens plane such that the cone closes at the source redshift. 
	
	The addition of halos along the line of sight and specifying a flat cosmology introduces an artificial focusing of light rays. To counteract this effect, we add negative convergence sheets along the line of sight computed with respect to the mean mass in dark matter halos we render \citep[see][]{Birrer++17b}.
	
	\subsection{The subhalo mass function of the main deflector}
	\label{ssec:submfunc}
	We parameterize the subhalo mass function in terms of a projected number density per unit area $\Sigma_{\rm{sub}}$. In principle, the abundance and spatial distribution of substructure depends on the total mass of the parent dark matter halo and redshift \citep{Gao++11,Han++16}, and tidal stripping, which can dramatically reduce the subhalo content of galactic halos \citep{DespVeg16,Han++16,GK++17,JiangvdB17,Richings++18}. We may therefore write the subhalo mass function as
	\begin{equation}
		\label{eqn:submfunc}
		\frac{d^2N}{dmdA} = \frac{\Sigma_{\rm{sub}}}{m_0} \left(\frac{m}{m_0}\right)^{-\alpha} F\left(M\right) H\left(z\right)
	\end{equation}
	where $F$ and $H$ encode dependence on the parent halo mass $M$ and redshift, respectively.  
	
	We render subhalos out to a maximum projected radius of $R_{\rm{max}} = 3 R_{\rm{Ein}}$, and render the subhalo z-coordinates in three dimensions out to the virial radius of the parent halo. In the semi-cylindrical volume defined by the viral radius and $R_{\rm{max}}$, we assume the spatial distribution of subhalos follows the mass profile of the parent NFW halo outside $r_{\rm{3D}} = 0.5 R_{\rm{s}}$, where $R_s$ is the scale radius of the parent halo, and assume the spatial distribution (per unit volume) is constant inside $0.5 R_{\rm{s}}$. This reflects the impact of tidal stripping, which tends to preferentially destroy subhalos in the central regions of halos \citep{JiangvdB17}. This procedure sets the distribution of subhalo z-coordinates, which affects the truncation of subhalos through Equation \ref{eqn:truncrad2}. When we render halo populations from this mass function and the line of sight halo mass function, we draw from a Poisson distribution with mean $\langle N \rangle$ obtained by normalizing and integrating Equation \ref{eqn:submfunc} (see Section \ref{sec:simsetup}).
	
	\subsection{Modeling free-streaming effects in WDM}
	\label{ssec:wdmmodel}
	
	Diffusion of dark matter particles in the early universe suppresses small scale power in the matter power spectrum below a characteristic `free-streaming length' that depends on the WDM particle mass and formation mechanism. For a more detailed review of WDM theory, see \citet{Benson++13,Schneider++13}.
	
	We parameterize free-streaming effects on the mass function through the half-mode mass $m_{\rm{hm}}$, defined with respect to the length scale where the WDM transfer function is damped with respect to the CDM transfer function by one-half. In WDM models, the number of halos below $m_{\rm{hm}}$ is strongly suppressed with respect to CDM. We adopt the functional form for this effect given by \citet{Lovell++14}
	\begin{equation}
		\label{eqn:wdmmassfunc}
		\frac{dN_{\rm{wdm}}}{dm} = \frac{dN_{\rm{cdm}}}{dm} \left(1+\frac{m_{\rm{hm}}}{m}\right)^{-1.3}.
	\end{equation}
	We note that other parameterizations for the turnover in the mass function differ slightly from Equation \ref{eqn:wdmmassfunc} \citep[see ][]{Schneider++12,Benson++13}. For instance, the WDM mass functions by \citet{Benson++13} exhibit a harder turnover than the parameterization in Equation \ref{eqn:wdmmassfunc} due to physical effects, namely, the presence of dark matter velocity dispersion at early times. Other (non-physical) variables, including the different algorithms for identifying and assigning mass to halos, and the choice of window function used to compute the matter power spectrum, can yield different mass functions for the same dark matter model. We do not explicitly address these complications in this work. Finally, we note that the effects of dark matter free-streaming may be enhanced at high redshift, suppressing halo counts relative to CDM more than that predicted by Equation \ref{eqn:wdmmassfunc}. This would increase the disparity between CDM and WDM on small scales, which would result in stronger constraints on $\mhm$ than those we project in this work. However, lacking a clear prediction for the redshift evolution of the WDM mass function, we do not model the effect in our forecasts. 
	
	\begin{figure*}
		\includegraphics[clip,trim=1cm 8.2cm 0.5cm
		5cm,width=.96\textwidth,keepaspectratio]{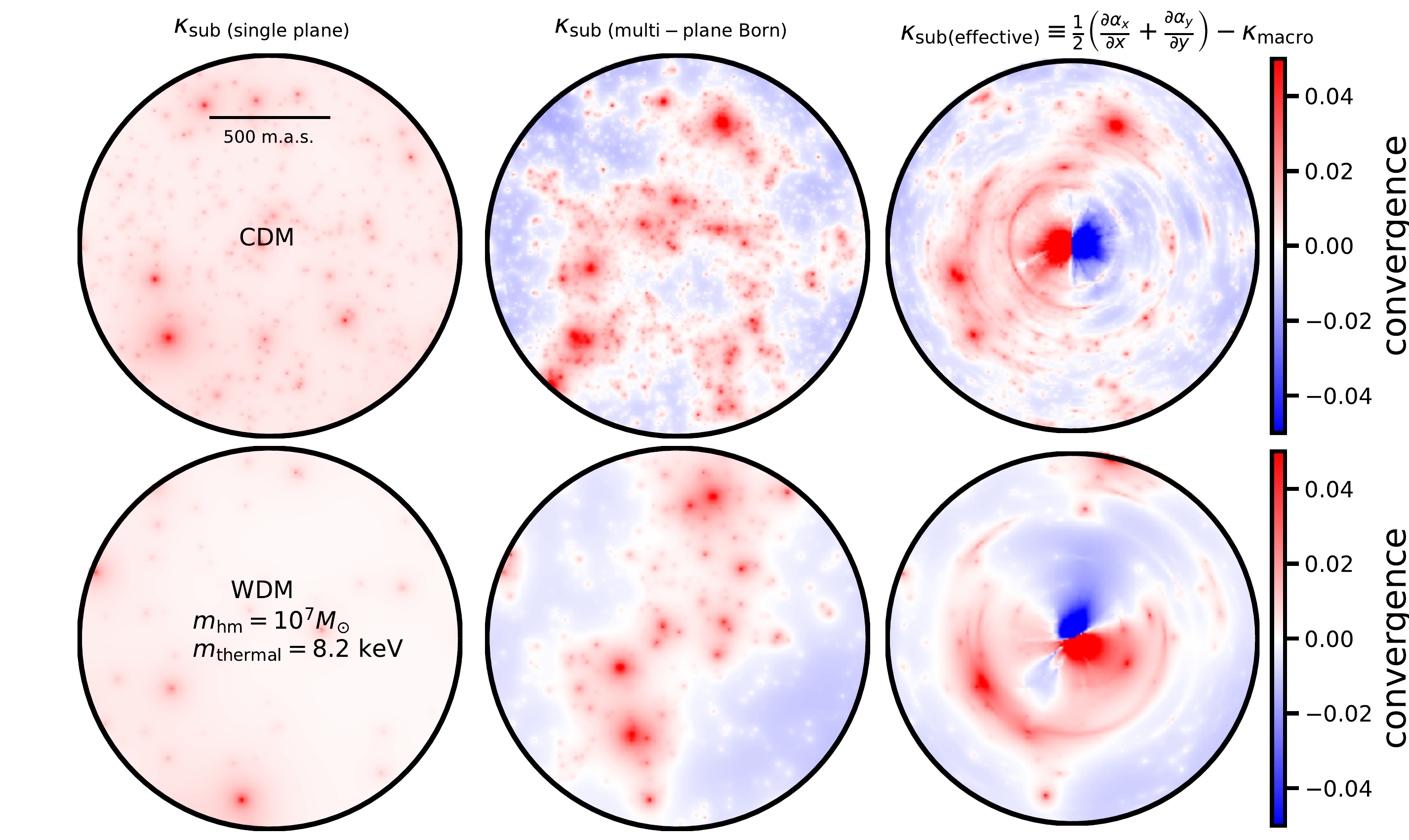}
		\caption{\label{fig:kappa_effective} A visualization of the mass distributions that affect observables in single-plane and multi-plane lensing. The top and bottom rows show a single realization of CDM and WDM structure, respectively. {\bf{Left:}} The convergence map from subhalos of the main deflector only, with $\Sigma_{\rm{sub}} = 0.012 \rm{kpc^{-2}}$, which corresponds to a projected mass fraction in substructure at the Einstein radius of $1\%$ at $z = 0.5$. {\bf{Center:}} The full line of sight realization viewed in projection. Computing deflection angles with respect to these mass distribution effectively employs the Born approximation, in which the deflection angles from halos at different redshifts are computed by assuming light travels along an unperturbed path. There are blue regions with negative mass due to the inclusion of negative convergence sheets at each lens plane (see discussion in Section 2.2). {\bf{Right:}} The \textit{effective multi-plane convergence} for these realizations. The deflection angles corresponding to these convergence maps, after subtracting off the convergence from the main deflector, include the non-linear effects present in multi-plane lensing not captured by the Born approximation (see Appendix \ref{app:A}).}
	\end{figure*}	
	\begin{figure}
		\includegraphics[clip,trim=0cm 0cm 0cm
		0cm,width=.48\textwidth,keepaspectratio]{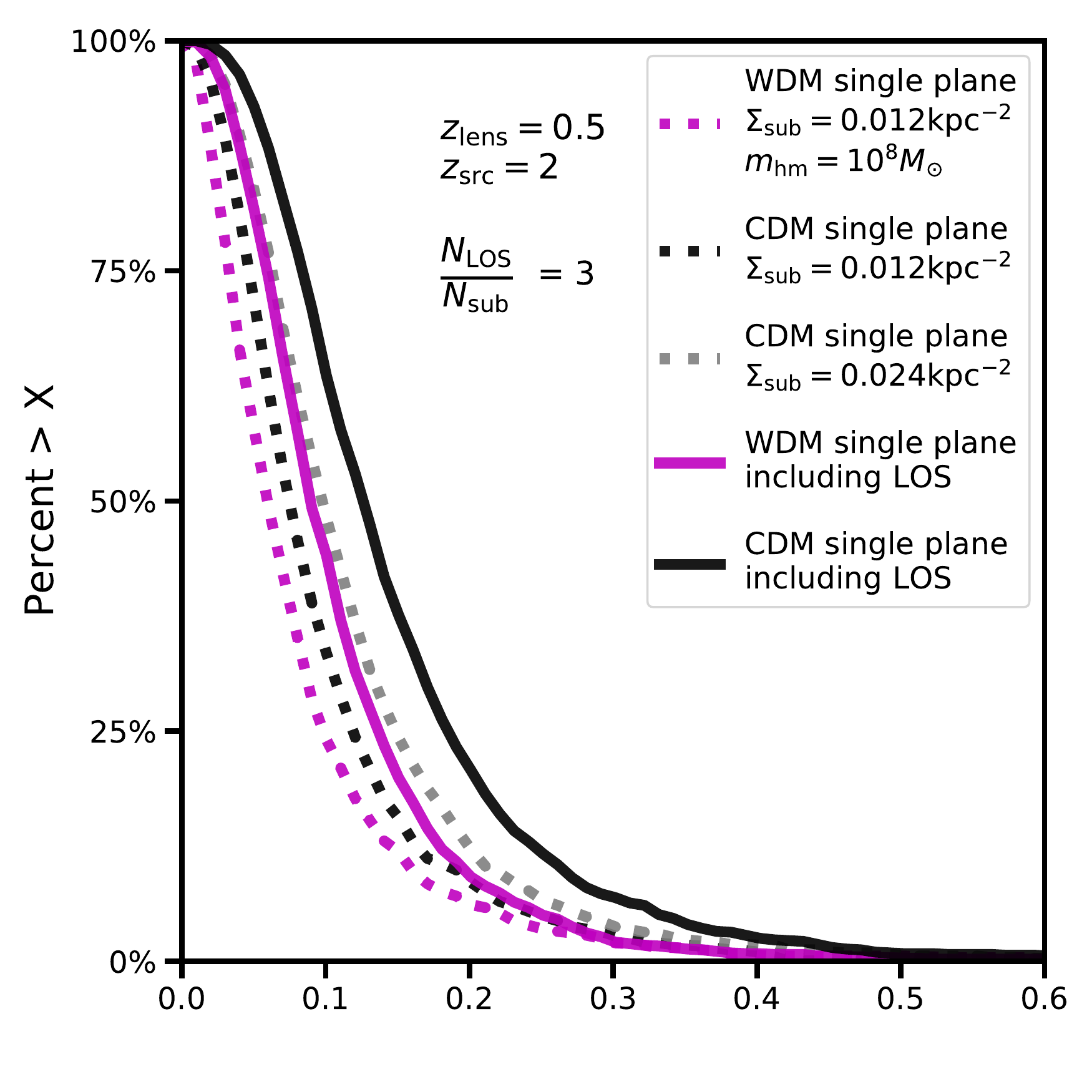}
		\includegraphics[clip,trim=0cm 0cm 0cm
		0cm,width=.48\textwidth,keepaspectratio]{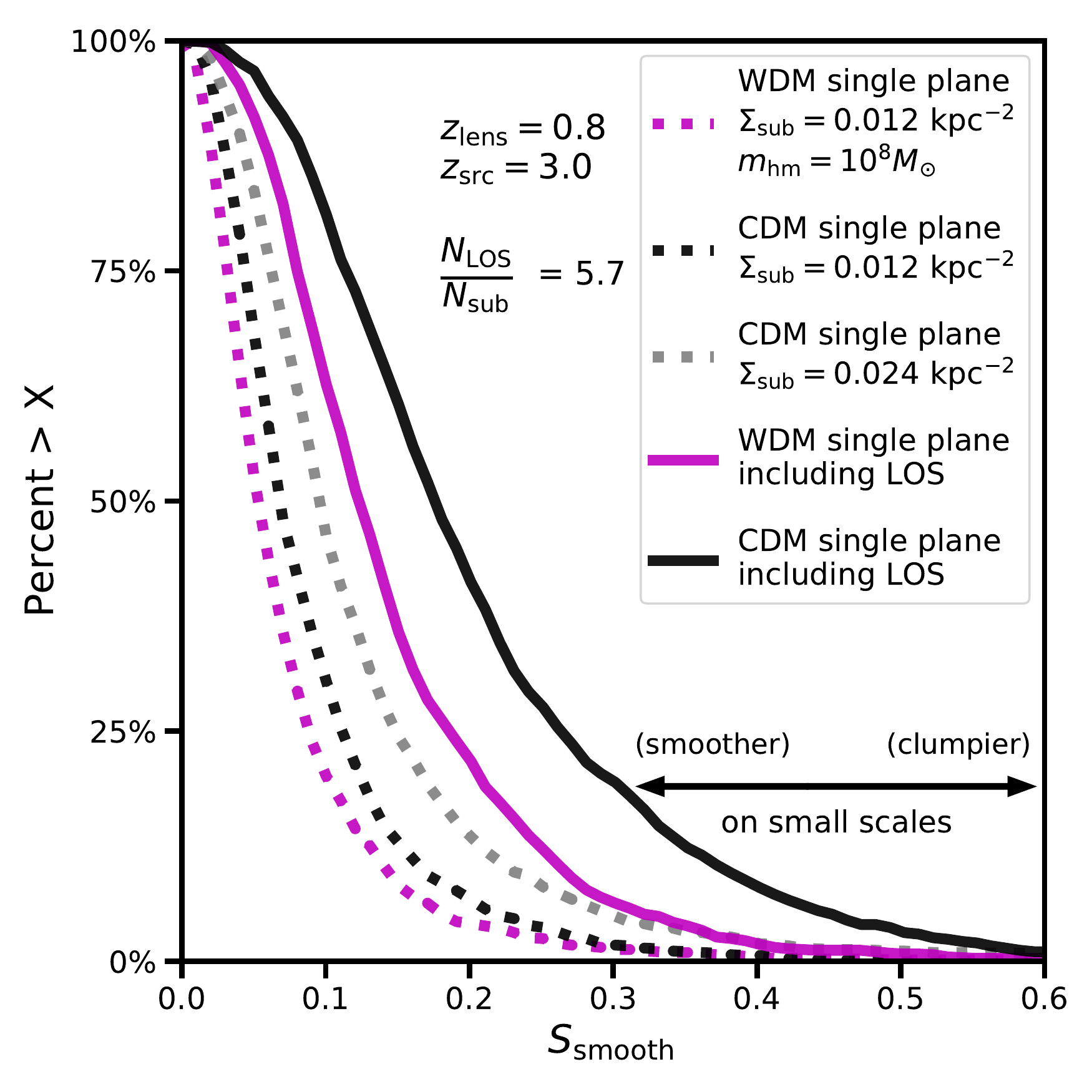}
		\caption{\label{fig:fluxdistributions} Distributions of the summary statistic in Equation \ref{eqn:summary_statsmooth} for different dark matter mass functions, and lens and source redshifts. Dotted curves represent realizations of main deflector subhalos only, while the solid curves include both subhalos and line of sight halos. Black and grey curves denote CDM mass functions with normalizations $\Sigma_{\rm{sub}} = 0.012 \rm{kpc^{-2}}$ and $0.024 \rm{kpc^{-2}}$, respectively, while magenta curves correspond to WDM mass functions with $\Sigma_{\rm{sub}} = 0.012 \rm{kpc^{-2}}$ and $\mhm = 10^{8} \msun$. Mass functions with more small scale structure produce more frequent flux ratio anomalies with respect to smooth lens models, which results in longer tails in the cumulative distribution of these statistics. The boost in the frequency and magnitude of flux ratio anomalies is much stronger for configurations with higher lens/source redshifts.}
	\end{figure}
	
	Thermally produced dark matter particles (thermal relics), assuming they comprise the entirety of the dark matter, admit a one-to-one mapping between the half-mode mass $m_{\rm{hm}}$ and the mass of the dark matter particle $m_{\rm{DM}}$. To translate between these two quantities, we use the scaling $m_{\rm{hm}} \sim m_{\rm{DM}}^{-3.33}$ \citep[see][]{Schneider++12}, and normalize this relation using the $2\times 10^8 \msun h^{-1} \sim 3.3 \ \rm{keV}$ constraint from the Lyman-$\alpha$ forrest \citep{Viel13}. This yields
	
	\begin{equation}
		\label{eqn:masskev}
		m_{\rm{hm}}\left(m\right) = 10^{10} \left(\frac{m_{\rm{DM}}}{\rm{1keV}}\right)^{-3.33} \msun h^{-1}.
	\end{equation}
	
	In addition to a suppressed mass function below the free streaming scale, free streaming alters the concentration-mass relation of WDM halos \citep{Schneider++12,Maccio++13,Bose++16,Ludlow++16}. We model this suppression using the parameterization given by \citep{Bose++16} 
	\begin{equation}
		\label{eqn:cmrelation}
		\frac{c_{\rm{wdm}}\left(m, z\right)}{c_{\rm{cdm}}\left(m, z\right)} =  \left(1+z\right) ^{\beta\left(z\right)} \left(1+60\frac{m_{\rm{hm}}}{m}\right)^{-0.17}.
	\end{equation}
	with $\beta\left(z\right) = 0.026z - 0.04$.\footnote{We remind the reader that we use the $c_{\rm{cdm}}\left(m,z\right)$ relation presented in \citet{DiemerJoyce18}.} We plot the subhalo mass function and the mass concentration relation in Figure \ref{fig:mcrelation}. Due to the factor of 60 in Equation \ref{eqn:cmrelation}, the effect on halo concentrations affects the central densities of objects with masses significantly above the half-mode mass.  
	
	\section{Effect of line of sight structure on image flux ratios}
	\label{sec:multiplanelensing}
	In order to constrain different dark matter models, we must accurately predict image flux ratios in the presence of perturbing dark matter halos in the main lens plane and along the line of sight. To this end, in this exploratory section we investigate the effect of halos at multiple redshifts on flux ratio observables. First, we present visualizations of the non-linear effects present in multi-plane lensing by defining an effective single plane mass distribution for a multi-plane lens system. We then quantify the signal in flux ratios from line of sight structures using a summary statistic, and compare the contributions from subhalos in the main deflector to the signal from line of sight objects for lenses at different redshifts.
	
	\subsection{Multi-plane lensing}
	\label{ssec:multplane}
	As photons traverse the cosmos from a background source to an observer, they experience numerous deflections by dark matter halos along the line of sight. One formulation of the equation describing these deflections and the path of deflected light rays is given by \citep{Schnedier1997}
	\begin{equation}
		\label{eqn:raytracing}
		\boldsymbol{\beta_S} = \boldsymbol{\theta} - \frac{1}{D_s} \sum_{n=1}^{S-1} D_{ns}{\boldsymbol{\alpha_n}} \left(D_n \boldsymbol{\beta_n}\right).
	\end{equation}
	where $\boldsymbol{\beta_S}$ and $\boldsymbol{\theta}$ denote angular coordinates in the source plane and on the sky, respectively, and where $D_n$ and $D_{ns}$ denote angular diameter distances to the $n$th lens plane, and between the $n$th lens plane and the source plane. 
	
	In the case of a single lens plane, the deflection field from multiple halos is a linear superposition of the deflections from each individual halo. In the case of multiple lens planes, however, Equation \ref{eqn:raytracing} becomes a recursive equation for the $\boldsymbol{\beta_n}$, coupling the deflections from halos at different redshifts. Equation \ref{eqn:raytracing} describes a physical process akin to looking through a magnifying glass through the lens of another magnifying glass (or in the case of substructure lensing, through thousands of other magnifying glasses). For additional details on multi-plane lensing, see \citet{Schneider++92}. 
	
	The number of halos along the line of sight often outnumber main lens plane subhalos, to a degree that depends on the lens and source redshifts, and the normalization of the subhalo mass function. However, number counts do not accurately reflect the effects of these line of sight objects on lensing observables. First, the geometry defined by the lens and source redshifts results in different lensing efficiencies for halos at different redshifts. Second, the coupling between deflections by halos at different redshifts results in non-linear effects that impact the deflection angles. 
	
	To glean some physical intuition of the lensing effects at play in a multi-plane system, we adopt a definition of the lensing surface mass density for multi-plane systems that encodes redshift-dependent lensing efficiency, and non-linear coupling between different lens planes. We define $\kappa_{\rm{effective}}$, the \textit{effective multi-plane convergence}, as
	
	\begin{equation}
		\label{eqn:kappasubeff}
		\kappa_{\rm{(effective)}}  \equiv \frac{1}{2} \ \div{ \boldsymbol{\alpha}}
	\end{equation}
	where $\boldsymbol{\alpha}$ is the deflection field of the lens system, or the mapping from a coordinate on the sky to a position in source plane through multi-plane ray-tracing.
	
	This definition expresses the convergence of a multi-plane realization in terms of deflections angles $\left(\alpha_x, \alpha_y\right)$ rather than a lensing potential, but is equivalent to the usual definition of convergence in the case of a single lens plane. \footnote{Convergence is equivalent to the projected surface mass density in units of the critical density for lensing $\Sigma_{\rm{crit}} = \frac{c^2}{4 \pi G}\frac{D_s}{D_{ds} D_d}$ in single plane lensing, where subscripts $d$ and $s$ denote the lens and source redshifts. For multiple lens planes, we express $\kappa$ as a vector-field derived quantity.} We compute these deflection angles by ray-tracing through the line of sight according to Equation \ref{eqn:raytracing}. To obtain an effective substructure convergence $\kappa_{\rm{sub(effective)}}$, we simply subtract the convergence profile of the main deflector $\kappa_{\rm{macro}}$ (the macromodel), from the full $\kappa_{\rm{(effective)}}$. 
	
	The definition of $\kappa$ in Equation \ref{eqn:kappasubeff} permits a comparison between single plane and multi-plane `convergence' maps. For illustrative purposes, in Figure \ref{fig:kappa_effective}, we render a full multi-plane realization of NFW halos between $10^{5.7}$ and $10^{10} \msun$, for a CDM and WDM scenario. The far left panels show only the single-plane realizations of the subhalo mass function, as would be present in a typical strong lens halo. The central panels show the single plane realizations plus the a full line of sight realization viewed in projection, with coupling between the multiple lens planes turned off. The lensing properties of this convergence map correspond to adopting the Born approximation in lensing, in which lensing quantities are computed by assuming the light rays follow unperturbed paths through the lens planes in front of and behind the main deflector. The far right panels show the \textit{effective multi-plane convergence} for these realizations. In Appendix \ref{app:B}, we compare flux ratios computed with the Born approximation to those computed with full ray-tracing, and find the two approaches yield significantly different observables. 
	
	Comparing the mass distribution in the far left panels with those on the far right suggests the inclusion of line of sight objects will dramatically affect the statistics of flux ratio distributions in strong lenses caused by small scale density fluctuations in the projected mass density. In the following sections, we will show that this is indeed the case.
	
	\subsection{Flux ratio statistics with line of sight halos}
	\label{ssec:lostats}
	We perform a simple experiment to build intuition for the impact of line of sight halos on flux ratio observables. First, we compute a set of image positions $\boldsymbol{x}$ and flux ratios $\boldsymbol{f_{\rm{reference}}}$ for a smooth lens mass distribution, which for simplicity we model as en elliptical isothermal-ellipsoid with external shear (SIE+Shear). Next, given a dark matter model with fixed $\Sigma_{\rm{sub}}$ and $\mhm$ (with $\delta_{\rm{los}} = 1$ and a background source size of 40 pc FWHM), we render 1,000 realizations of halos this model from Equations \ref{eqn:losmfunc} and \ref{eqn:submfunc}. For each of these realizations, we optimize a smooth model to fit the image positions, and compute the model flux ratios $\boldsymbol{f^{\prime}}$ with respect to this optimized lens model. We then compute the summary statistic\footnote{The summation $i$ runs over the three flux ratios derived from the four image fluxes.}
	\begin{equation}
		\label{eqn:summary_statsmooth}
		S_{\rm{smooth}} \left(\boldsymbol{f^{\prime}},\boldsymbol{f_{\rm{reference}}} \right) \equiv \sqrt{\sum_{i=1}^{3} \left({f}^{\prime}_{i} - f_{\rm{reference(i)}} \right)^2}.
	\end{equation}
	
	The statistic $S_{\rm{smooth}}$ encodes the amount of flux ratio anomaly with respect to a smooth lens model induced by the presence of dark matter halos. In principle, the distributions of this statistic depend on the reference smooth lens model used to compute $\boldsymbol{f_{\rm{reference}}}$, but since we construct these distributions merely for visualization purposes the choice of smooth model is not crucial.  These complications notwithstanding, we note that the SIE+Shear profile used to compute $S_{\rm{smooth}}$ reasonably describes the large-scale mass profile of a typical deflector \citep{Auger++10,Gilman++17}. 
	
	Figure \ref{fig:fluxdistributions} shows distributions of $S_{\rm{smooth}}$ for different lens (source) redshifts of 0.5 (2) and 0.8 (3) with different dark matter models. The addition of line of sight halos increases the frequency of a flux ratio anomaly with respect to a smooth lens model, and the boost is substantially higher for configurations with higher lens and source redshifts. The inclusion of line of sight structure also increases the difference in relative amplitudes between the CDM and WDM (solid black and magenta curves) relative to models with lens plane subhalos only. Finally, the distribution of summary statistics for a CDM mass function with a high normalization (grey dotted curve) resembles the statistics produced in a WDM model with a lower value of $\Sigma_{\rm{sub}}$. This reflects a degeneracy between the amplitude of the subhalo mass function in the main lens plane, and the turnover scale in the mass function. 
	
	In the next Section, we amend the definition of the summary statistic in Equation \ref{eqn:summary_statsmooth} slightly, replacing $\boldsymbol{f_{\rm{reference}}}$ with a set of observed fluxes from a strong lens $\boldsymbol{f_{\rm{obs}}}$. We write this new statistic $S_{\rm{lens}}$ as
	
	\begin{equation}
		\label{eqn:summary_stat}
		S_{\rm{lens}} \left(\boldsymbol{f^{\prime}},\boldsymbol{f_{\rm{obs}}} \right) \equiv \sqrt{\sum_{i=1}^{3} \left({f}^{\prime}_{i} - f_{\rm{obs(i)}} \right)^2}.
	\end{equation}
	
	Through the forward model, we will attempt to minimize this statistic by computing flux ratios $\boldsymbol{f^{\prime}}$ with different dark matter mass functions. The model flux ratios that minimize this statistic match the observed flux ratios at the particular image positions, and as such the model flux ratios minimizing the statistic satisfy the same correlations as those present in the data. In Appendix \ref{app:C}, we describe the implementation of a fast algorithm for lens model optimizations with many line of sight halos, which we use to compute the statistic in Equation \ref{eqn:summary_stat}. 
	
	\section{Simulations of substructure lensing: setup and methodology}
	\label{sec:simsetup}
	In this section, we describe the setup of simulations designed to project the constraining power of flux ratios on a WDM mass function. We first outline the physical assumptions imposed in the simulations, and the priors on the parameters sampled in the forward model. Next, we walk through the forward modeling procedure. The subsequent section describes our implementation of flux uncertainties, both from measurement errors and lens modeling. We then describe how, after accounting for uncertainty in the image fluxes, we construct posterior distributions for the model parameters. Finally, we describe the procedure for creating simulated datasets we will use to test this machinery and make forecasts.  
	
	\subsection{Physical assumptions and priors}
	\label{ssec:assumptionsandpriors}
	The methodology we present is flexible, and accommodates any parameterization for the quantities such as the subhalo mass function, line of sight halo mass function, main deflector mass profile, etc. However, for the purpose of making forecast statements and presenting the methodology, we make several simplifying assumptions regarding the implementation of dark matter physics, mass models, and lensing quantities. 
	
	\subsubsection{The subhalo mass function}
	First, we do not marginalize over the mass, concentration, or ellipticity of the host dark matter halo. We assume a halo mass of $10^{13}\msun$, which is typical for a lensing galaxy \citep{Gavazzi++07}, when distributing halos spatially and evaluating the two-halo term in Equation \ref{eqn:losmfunc}. We do not expect the ellipticity of the parent dark matter halo to affect the lens model predictions for image fluxes, since the ellipticity of the lensing galaxy and external shear dominate the quadrupole moment of the mass distribution \citep{Keeton++97}. We also ignore any redshift dependence in the subhalo mass function, although we evolve the line of sight halo mass function evolve with redshift. With these simplifications, the subhalo mass function in Equation \ref{eqn:submfunc} takes the form 
	\begin{equation}
		\label{eqn:submfunc2}
		\frac{d^2N_{(13)}}{dmdA} = \frac{\Sigma_{\rm{sub}}}{m_0} \left(\frac{m}{m_0}\right)^{-\alpha} 
	\end{equation}
	where the subscript (13) refers to the assumed halo mass of $10^{13} \msun$.  We assume $\alpha = 1.9$ \citep{Springel++08,Fiacconi++16}. 
	
	We derive a projected mass density in subhalos by integrating Equation \ref{eqn:submfunc2} over mass, and find values of $\Sigma_{\rm{sub}} \sim 0.01-0.02 \ \rm{kpc^{-2}}$ yield surface mass densities in substructure similar to those derived in simulations of early-type galaxy halos of $10^7 \msun \rm{kpc^{-2}}$ with a pivot mass of $m_0 = 10^{8} \msun$ \citep{Fiacconi++16}. This normalization in principle depends on the severity of tidal stripping, the host halo mass, the halo redshift, and the halo formation time. Rather than modeling all of these effects from first principles, we subsume them in the normalization $\Sigma_{\rm{sub}}$, and impose a wide (flat) prior on this parameter between $0-0.045 \rm{kpc^{-2}}$. \citet{Gilman++18} demonstrate that the mean normalization in the lens sample effectively scales the information content available per lens; we perform the same analysis in this work, examining how the constraints on dark matter respond to different values of $\Sigma_{\rm{sub}}$. 
	
	Given a detailed model for the redshift evolution and halo mass depedence of the normalization, as well as the effects of tidal stripping, a non-flat, more informative prior could be used. Since we lack this information, and since we subsume the halo mass dependence and redshift evolution into $\Sigma_{\rm{sub}}$, we assume minimum information and use a flat prior.
	\begin{table*}
		\centering
		\caption{Parameters sampled in the forward model}
		\label{tab:params}
		\begin{tabular}{lccr} 
			\hline
			parameter & definition & prior\\
			\hline 
			$\Sigma_{\rm{sub}} \left[\rm{kpc}^{-2}\right]$ & normalization of subhalo mass function (Equation \ref{eqn:submfunc2})&  uniform: $\left[0, 0.045\right] $\\&(rendered between $10^6-10^{10} \msun$) & \\
			\\
			$\mhm \left[M_{\odot}\right]$ & half-mode mass (Equations \ref{eqn:wdmmassfunc} and \ref{eqn:cmrelation})& log-uniform: $\left[4.8, 10\right] $ \\
			&$\propto$ to free streaming length and thermal relic mass $m_{\rm{DM}}$ &\\
			\\
			$\delta_{\rm{los}}$ & rescaling factor for the line of sight Sheth-Tormen & uniform: $\left[0.7, 1.3\right]$ \\
			&mass function (Equation \ref{eqn:losmfunc}, rendered between $10^6-10^{10} \msun$)&\\
			\\
			$\sigma_{\rm{src}} \left[\rm{pc}\right]$ & source size& uniform: $\left[25, 50\right]$\\
			& parameterized as FWHM of a Gaussian &\\
			\\
			$\gamma_{\rm{macro}}$ & logarithmic slope of main deflector mass model  & uniform: $\left[2, 2.2\right]$\\
			\\
			$\delta_{xy} \left[\rm{m.a.s.}\right]$ & image position uncertainties& $\mathcal{N}\left(0, 3\right)$\\
			\hline		
			
		\end{tabular}
	\end{table*}
	\subsubsection{Free streaming in WDM}
	Regarding the implementation of WDM mass functions, we assume that the parameterization of the mass function turnover near $\mhm$ (Equation \ref{eqn:wdmmassfunc}) applies to both halos along the line of sight, and for subhalos in the main lens halo. As we vary the half-mode mass $\mhm$ between $10^{4.8} - 10^{10} \msun$, none of the models considered are truly `cold' in the sense of GeV-scale WIMPS with free-streaming masses of order an Earth mass. However, provided $\mhm << m_{\rm{low}} = 10^6 \msun$, the halo populations rendered result in the same observables as those in a CDM universe. \footnote{This is only true if the signal in flux ratio saturates at $m_{\rm{low}}$, otherwise we would miss part of the signal from halos with mass $<m_{\rm{low}}$. We verify that halos of mass below $10^6 \msun$ do not significantly affect the flux ratio signal for the background source sizes 25-50 pc.} We therefore interpret inferences that favor models with $\mhm <  10^6 \msun$ as consistent with CDM, even though the true half-mode mass may be in fact be much lower than the value we recover. Finally, while we implement scatter and redshift dependence in the mass concentration relation in Equation \ref{eqn:cmrelation}, we do not marginalize over the parameters describing the turnover for WDM models. 
	
	\subsubsection{Halo and subhalo mass range}
	We render subhalos and line of sight halos in the mass range $10^6 - 10^{10} \msun$. We choose the lower bound by reducing the smallest rendered halo mass until the distributions of $S_{\rm{smooth}}$ (like those in Figure \ref{fig:fluxdistributions}) become insensitive to lower masses (see also footnote 6). On the other hand, halos more massive than the upper bound of $10^{10} \msun$ would likely host stars and be visible, allowing them to be directly included in the main lens model \citep[e.g.][]{Birrer++19}.
	
	\subsubsection{Scaling of the LOS halo mass function}
	We vary the rescaling parameter for the line of sight halo mass function between 0.7 and 1.3. This accounts for theoretical uncertainties in the prediction of the halo mass function, which is typically at the $10-30 \%$ level \citep{Despali++16}. This term also accounts for variance in the average density along the line of sight to strong lenses. This parameter is not meant to account for correlated structure near the main lens plane, which we model through the two-halo term $\xi_{\rm{2halo}}$.
	
	\subsubsection{The background source size}
	The background source size enters the forward model because the perturbation to image magnifications depends on the source size relative to the deflection angle of a perturber \citep{DoblerKeeton02}. Upper limits on the size of the narrow-line region from \citep{Nierenberg++17} correspond to physical sizes of $\sim 50 \rm{pc}$, which agrees with the surface brightness profiles seen in low redshift AGN \citep{MullerSanchez++11}. We therefore allow the source size to vary between 25 and 50 pc. While in this work we forward model source sizes appropriate for narrow-line emission, the method we present can accommodate flux ratios measured from any band provided it is free from contamination from micro-lensing, including mid-infrared bands \citep{Minezaki++09,MacLeod++13}. 
	
	\subsubsection{The main deflector}
	We model the main deflector as a power-law ellipsoid plus external shear. This is a generalization of the widely applied, physically motivated \citep[e.g.][]{Treu++06} singular isothermal sphere (SIE) profile used to model lensing galaxies. Studies of early-type deflectors find mass profiles $\rho\left(r\right) \sim r^{-\gamma_{\rm{macro}}}$ modestly steeper than $r ^{-2}$  \citep{Treu++09,Auger++10,Shankar++17}, so we allow the power-law profile $\gamma_{\rm{macro}}$ to vary between 2 and 2.2. We assume deflectors with complex morphologies, including features like stellar disks, have been identified and removed from our sample, and describe residual baryonic effects by adding perturbations to the forward model image fluxes, a process we describe in Section \ref{ssec:fluxdelta}. We marginalize over uncertainties in image positions by rendering Gaussian astrometric uncertainties of $\pm 3$ m.a.s. in the forward model.  
	
	\subsubsection{Summary}
	We point out that many of the simplifying assumptions we impose in our forecasts effectively ignore relevant information that could be used to inform a prior. For example, the velocity dispersion of the lensing galaxy could inform a prior on the halo mass and the normalization $\Sigma_{\rm{sub}}$, and possibly the macromodel profile $\gamma_{\rm{macro}}$. Since $\Sigma_{\rm{sub}}$ is somewhat correlated with $\mhm$ (see Section \ref{sec:results}), this could improve constraints on the free-streaming length of the dark matter. Similarly, modeling redshift dependence in the normalization of the subhalo mass function could break the covariance between $\Sigma_{\rm{sub}}$ and $\delta_{\rm{los}}$ (see Section \ref{sec:results}). This information would therefore improve the precision on the inferred dark matter properties, and it is possible that we overestimate uncertainties by omitting it.    
	
	\subsection{Forward modeling procedure}
	\label{ssec:forwardmodeling}
	
	To constrain the halo mass function, we adopt a forward modeling approach. This consists of generating mock data sets by simulating the physical processes that affect lensing observables, including the size of the background source, dark matter halos in the main lens halo and along the line of sight, the mass profile of the main deflector, and statistical measurement errors. This approach handles complicated degeneracies between model parameters - for example, between halo redshift and halo mass \citep[e.g.][]{Despali++18} - by building these features directly into the forward-generated data sets. In effect, we exchange the task of computing a complicated likelihood function with the challenge of simulating the relevant physics in strong lensing.
	
	This first step in the forward model is to sample all parameters from their respective prior probability densities, summarized in Table \ref{tab:params}. For convenience, for the $i$th realization, we denote the collection of the model parameters $\boldsymbol{M_i}$. Using the parameters describing the dark matter $\left(\Sigma_{\rm{sub}}, \delta_{\rm{los}}, \mhm \right)$, we render a the full population of line of sight halos and lens plane subhalos, as described in Section \ref{sec:rendering}. 
	
	Next, using the observed image positions \footnote{We add random statistical measurement errors of $\pm 3$ m.a.s. to the image positions for each realization.} and fluxes from a strong lens, we optimize a power-law plus external shear lens model with power law slope $\gamma_{\rm{macro}}$ to fit the observed image positions in the presence of the full population of dark matter halos, and ray-trace to compute the flux ratios with background source modeled as a Gaussian with a FWHM of $\sigma_{\rm{src}}$. While optimizing the macromodel to fit image positions, we allow the lens Einstein radius, centroid, ellipticity, ellipticity angle, shear, and shear angle to vary, while keeping the power-law slope $\gamma_{\rm{macro}}$ fixed for each optimization. If necessary, we may extend the forward modeling of $\gamma_{\rm{macro}}$ to additional mass profile parameters to add complexity in the lens macromodel. 
	
	At this stage, we have a set of observed flux ratios and a set of flux ratios simulated in the forward model. We use the model-predicted flux ratios $\boldsymbol{f^{\prime}}$ with the observed flux ratios $\boldsymbol{f}_{\rm{obs}}$ to compute the summary statistic in Equation \ref{eqn:summary_stat}, which we then assign to the set of parameters $\boldsymbol{M_i}$. We repeat this entire procedure 600,000 times for each lens (see the convergence test in Appendix \ref{app:D}). 
	
	\subsection{Accounting for uncertainty in image fluxes}
	\label{ssec:fluxdelta}
	We introduce uncertainties in the image fluxes by adding perturbations to the fluxes in the mock data, and by rendering these perturbations in the model fluxes. Explicitly, we modify each model-predicted image flux $f_i$ as
	\begin{equation}
		f_i \rightarrow f_i + \mathcal{N} \left(0, \delta \right).
	\end{equation} 
	The most straightforward interpretation of this procedure is the incorporation of statistical measurement errors. For reference, current measurements of narrow-line fluxes achieve precision of $3-6 \%$ \citep{Nierenberg++14,Nierenberg++17}. These perturbations also simulate the role of unknown sources of uncertainty, or simply those we do not explicitly model. For example, in cases where a more complex macromodel is required, the additional degrees of freedom that must be marginalized over result in a larger variation in image fluxes at fixed image positions, which effectively introduces an additional source of flux uncertainty. 
	
	We will explicitly consider flux perturbations of $2\%$, $4\%$, $6\%$, and $8\%$. The intermediate values of $4\%$ and $6\%$ represent current measurement precision \citep{Nierenberg++17} and modeling uncertainties \citep{Gilman++17}. The $2\%$ value represents a best-case scenario with precise measurements --- perhaps with observations from future telescopes such as JWST --- and a sample of morphologically simple deflectors that do not require complex macromodels. The $8\%$ value corresponds to a scenario where the majority of the systems in the lens sample require marginalization over complex macromodels.  
	
	\subsection{Bayesian Inference}
	\label{ssec:forwardmodeling}
	To construct posterior probability densities for the parameters $\boldsymbol{M}$ listed in Table \ref{tab:params}, we rank the 600,000 $\boldsymbol{M_i}$ by their summary statistics, with those that minimize the statistic ranked highest. A subset of these models (we use the top 1,500) form a probability density $p^{\prime}\left(\boldsymbol{M} | \rm{data}\right)$, which becomes an increasingly good approximation of the true posterior distribution $p\left(\boldsymbol{M} | \rm{data}\right)$ as the number of forward model samples increases. This procedure falls in the category of Approximate Bayesian Computing methods (for a review, see \citep{Lintusaari++17}), and is widely applied to problems with intractable likelihood functions \citep{Akeret++15,Hahn++17,Birrer++17a,Davies++18}. We apply a kernel density estimator to the 1,500 sample that form $p^{\prime}\left(\boldsymbol{M} | \rm{data}\right)$, and multiply the resulting probability densities to obtain the final posterior. We test for convergence in this algorithm in Appendix \ref{app:D}. 
	
	We acknowledge that, formally, a marginalization of the macromodel, rather than an optimization of the macromodel, yields the desired posterior distribution of dark matter parameters. We avoid this computationally prohibitive step \footnote{This is computationally prohibitive because the vast majority of macromodel parameter configurations do not fit the image positions, and therefore consume computation time without contributing to the desired posterior distribution.} with two justifications: First, the volume of macromodel parameter space is typically tightly constrained by the requirement that the macromodel fit the image positions. For macromodels parameterized as power-law ellipsoids, the image fluxes do not vary significantly over this volume, and the variation in image fluxes induced by marginalizing over the macromodel is negligible compared to other sources of uncertainty \footnote{We test this by re-sampling a once-optimized macromodel around the peak of the likelihood, and computing the variation in image fluxes.} Second, we note that for each of the 600,000 realizations rendered in the forward model, each macromodel re-optimization is independent. Thus, over the course of many realizations, covariance between macromodel parameters and the parameters describing the dark matter content is reflected in the summary statistics. 
	
	\subsection{Creating simulated data sets}
	\label{ssec:simdata}
	
\begin{figure}
		\includegraphics[clip,trim=0cm 0cm 0cm
		1.5cm,width=.48\textwidth,keepaspectratio]{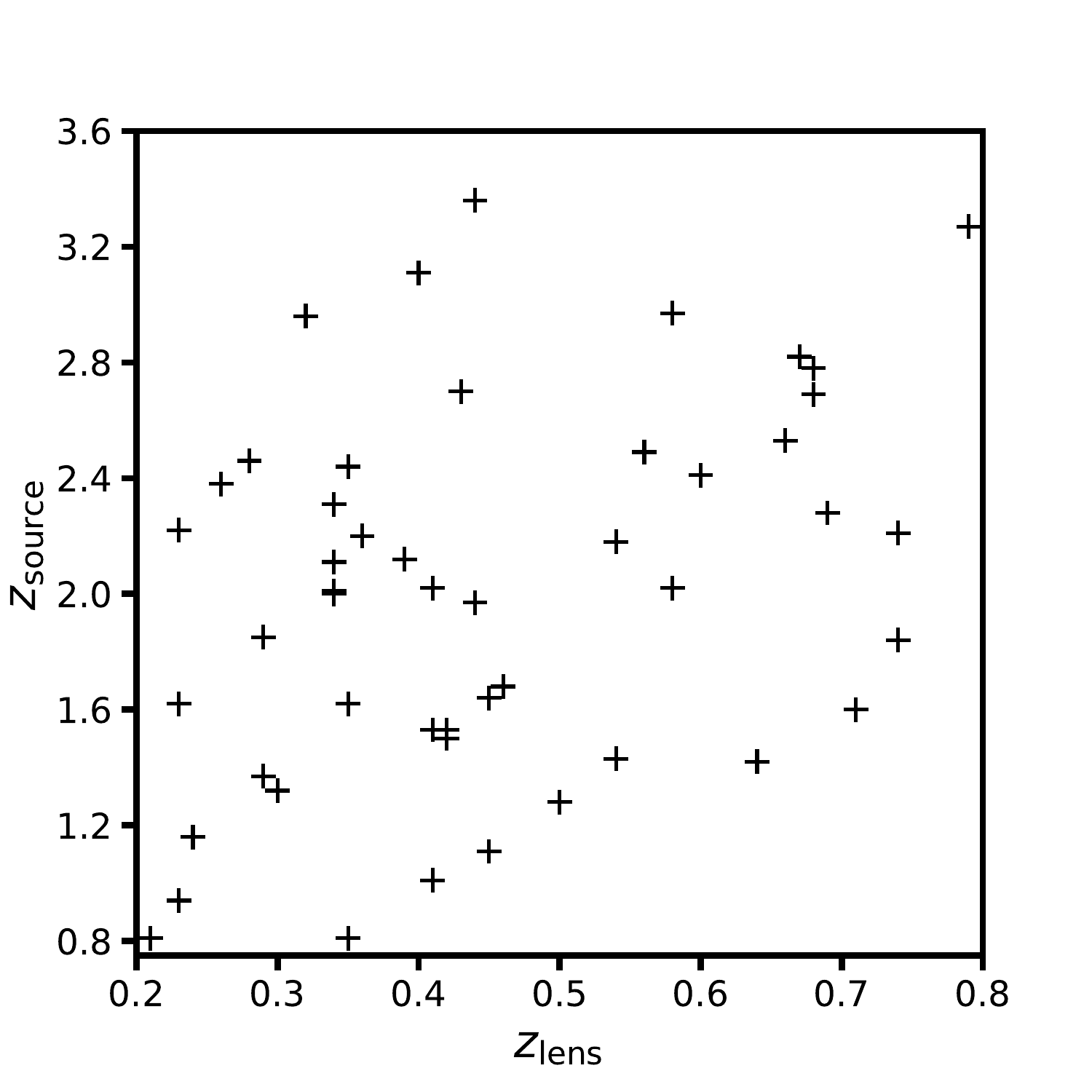}
		\caption{\label{fig:zlenszsrc} The lens and source redshifts for the 50 quads in our mock lens sample. We draw these parameters, along with the lens velocity dispersion, ellipticity, and shear from the distributions used by \citet{OguriMarshall10}. }
	\end{figure}

	To create mock data sets, we parameterize the lens macromodel as a power-law ellipsoid, and generate mock lenses by sampling the Einstein radii, ellipticity, and external shears, as well as lens and source redshifts, from the distributions of these quantities used by \citet{OguriMarshall10}. We plot the lens and source redshifts of the 50 quads in our mock lens sample in Figure \ref{fig:zlenszsrc}. We sample power law slopes drawn from a distribution centered at $2.05 \pm 0.04$, consistent with the morphological properties of the early-type galaxies that dominate the strong lensing cross section \citep{Auger++10,Shankar++17}. The background source is parameterized by a circular Gaussian with a FWHM, which we specify within the range $25-50$ pc, consistent with the upper limits on the size inferred by \citet{Nierenberg++17}, and comparable to the luminous extent of the narrow line region of quasars \citep{MullerSanchez++11}. 
	
	We choose background source positions to produce roughly equal numbers of cross, fold, and cusp image configurations. Cusp and fold configurations generally yield the strongest constraints on WDM properties (see Appendix B in \citet{Gilman++18}), and since the images in these types of quads have higher magnifications they may be more easily discovered. It is therefore possible that a real sample of quads would consist of more cusp and fold configurations than crosses, in which case the resulting constraints on WDM would be stronger than those obtained in this work.
	
	When generating the mock data sets, we add measurement errors to the image positions of $3$ m.a.s., and model statistical measurement errors by adding perturbations to the image fluxes, as described in Section \ref{ssec:fluxdelta}.
	
	\section{Simulations of substructure lensing: Results}
	\label{sec:results}
	\begin{figure*}
		\includegraphics[clip,trim=0cm 0cm 0cm
		0cm,width=.9\textwidth,keepaspectratio]{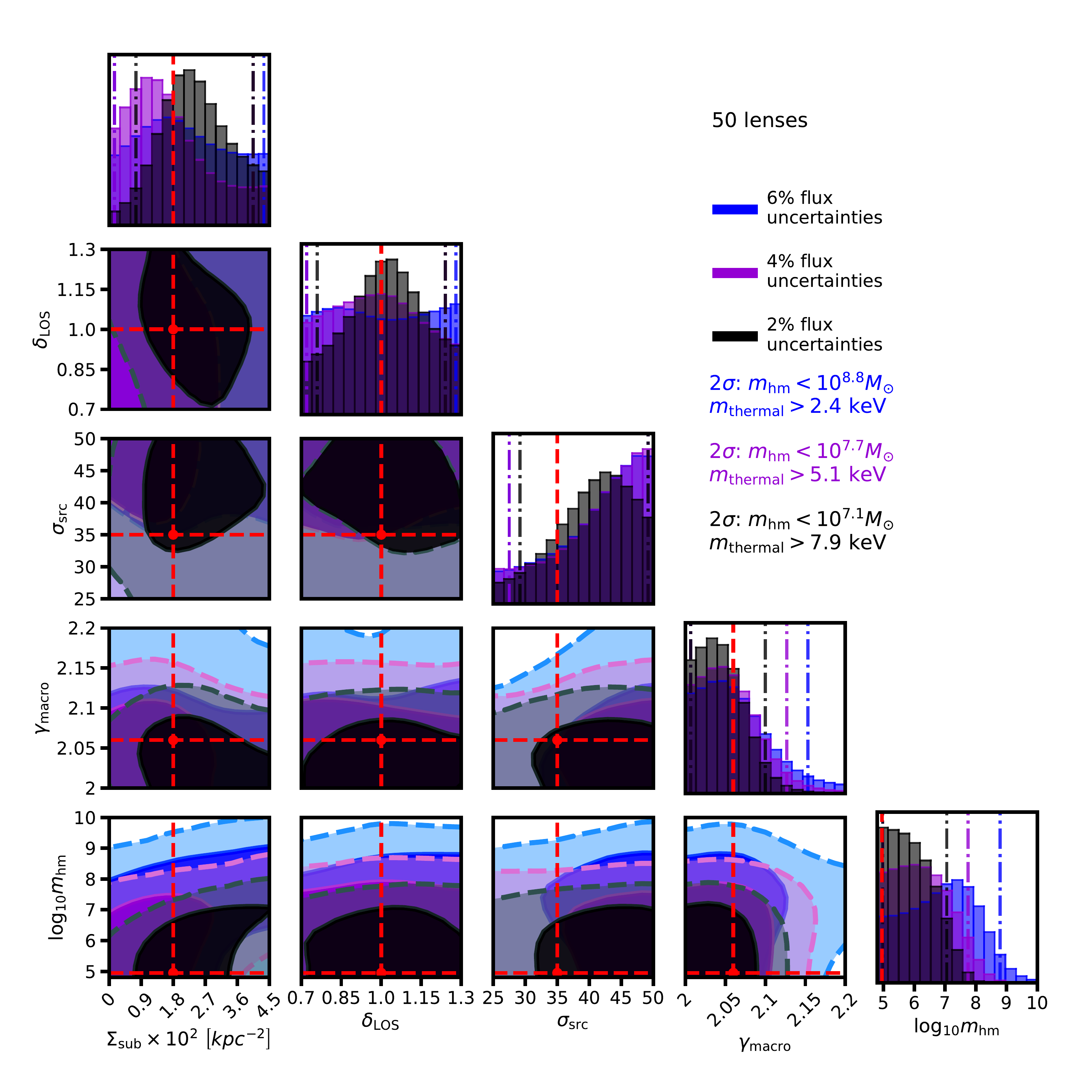}
		\caption{\label{fig:CDM_1} The posterior distributions resulting from the forward modeling analysis of a sample of 50 lenses, with flux uncertainties stemming from measurement errors and lens modeling controlled at the $2\%$, $4\%$, and $6\%$ level. Vertical bars in the marginal distribution indicate $2 \sigma$ confidence interval, while dashed (solid) lines in the panels denote $2 \sigma$ $\left(1 \sigma\right)$ contours. The marginalized constraints on $\mhm$ range from $10^{7.2} \msun$ for the case of $2 \%$ flux uncertainties, to $10^{8.8} \msun$ for uncertainties of $6 \%$.}
	\end{figure*}	
	\begin{figure*}
		\includegraphics[clip,trim=2cm 0cm 3.5cm
		0.25cm,width=.48\textwidth,keepaspectratio]{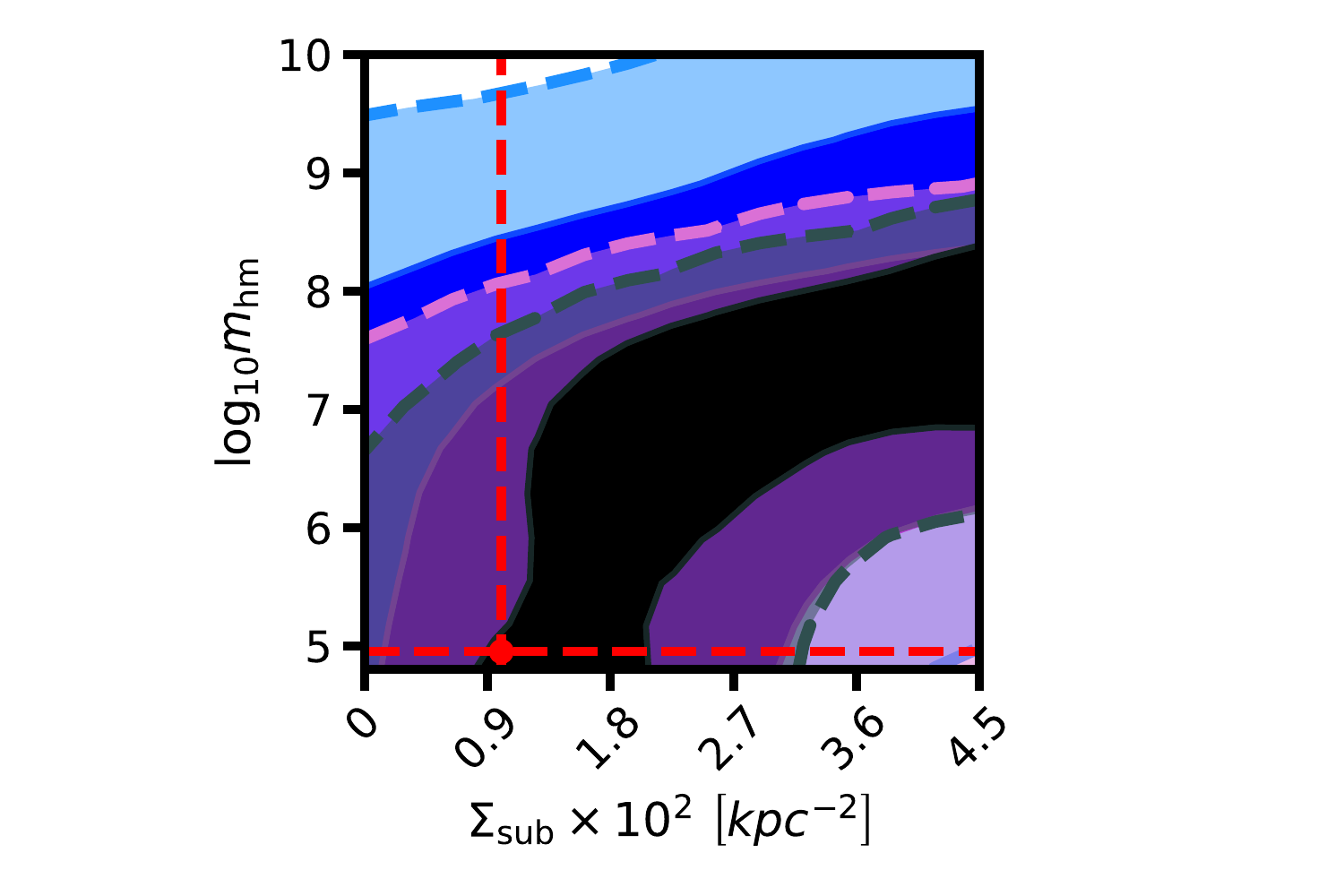}
		\includegraphics[clip,trim=2cm 0cm 3.5cm
		0.5cm,width=.48\textwidth,keepaspectratio]{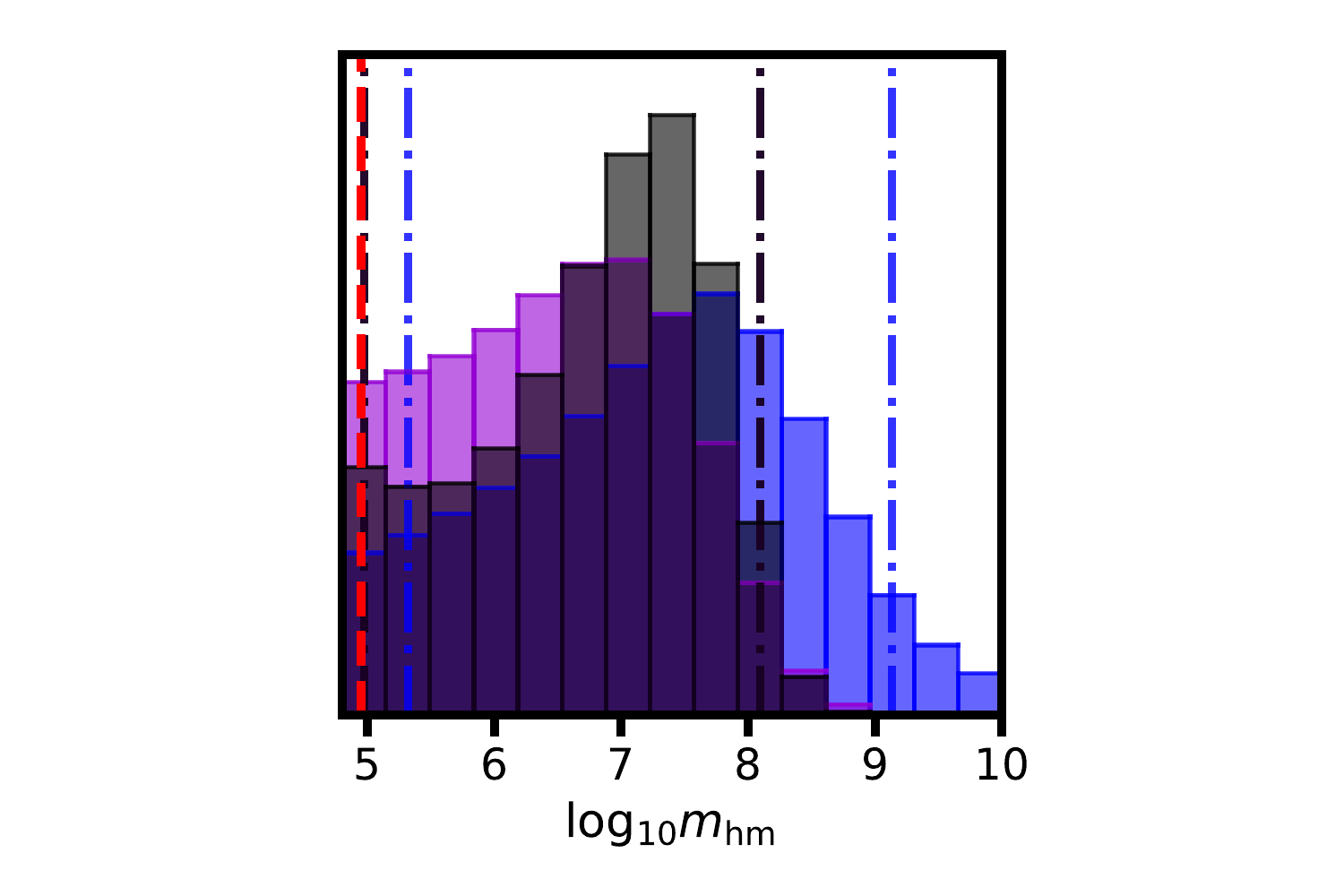}
		\caption{\label{fig:fluxuncertainties_CDM1} Inference on a CDM mass function with a normalization of the subhalo mass function $\Sigma_{\rm{sub}} = 0.01$, roughly half the value of the normalization assumed in Figure \ref{fig:CDM_1}. The color scheme is the same as in Figure \ref{fig:CDM_1}, with black, magenta, and blue representing flux uncertainties of $2\%$, $4\%$, and $6\%$, respectively. In this case, the marginalized constraints on $\mhm$ are $10^{9.1} \msun$, $10^{8.1} \msun$, and $10^{8.1} \msun$ for flux uncertainties of $6 \%$, $4\%$, and $2 \%$ (for the $4\%$ and $2\%$ flux uncertainties, the $2 \sigma$ confidence interval happen to be the same). These constraints are weaker by roughly an order of magnitude in mass over the bounds quoted in Figure \ref{fig:CDM_1}, which illustrates the role of the normalization of the subhalo mass function on the possible constrains on $\mhm$.}
	\end{figure*}	
	\begin{figure*}
		\includegraphics[clip,trim=2cm 0cm 3.5cm
		0.25cm,width=.48\textwidth,keepaspectratio]{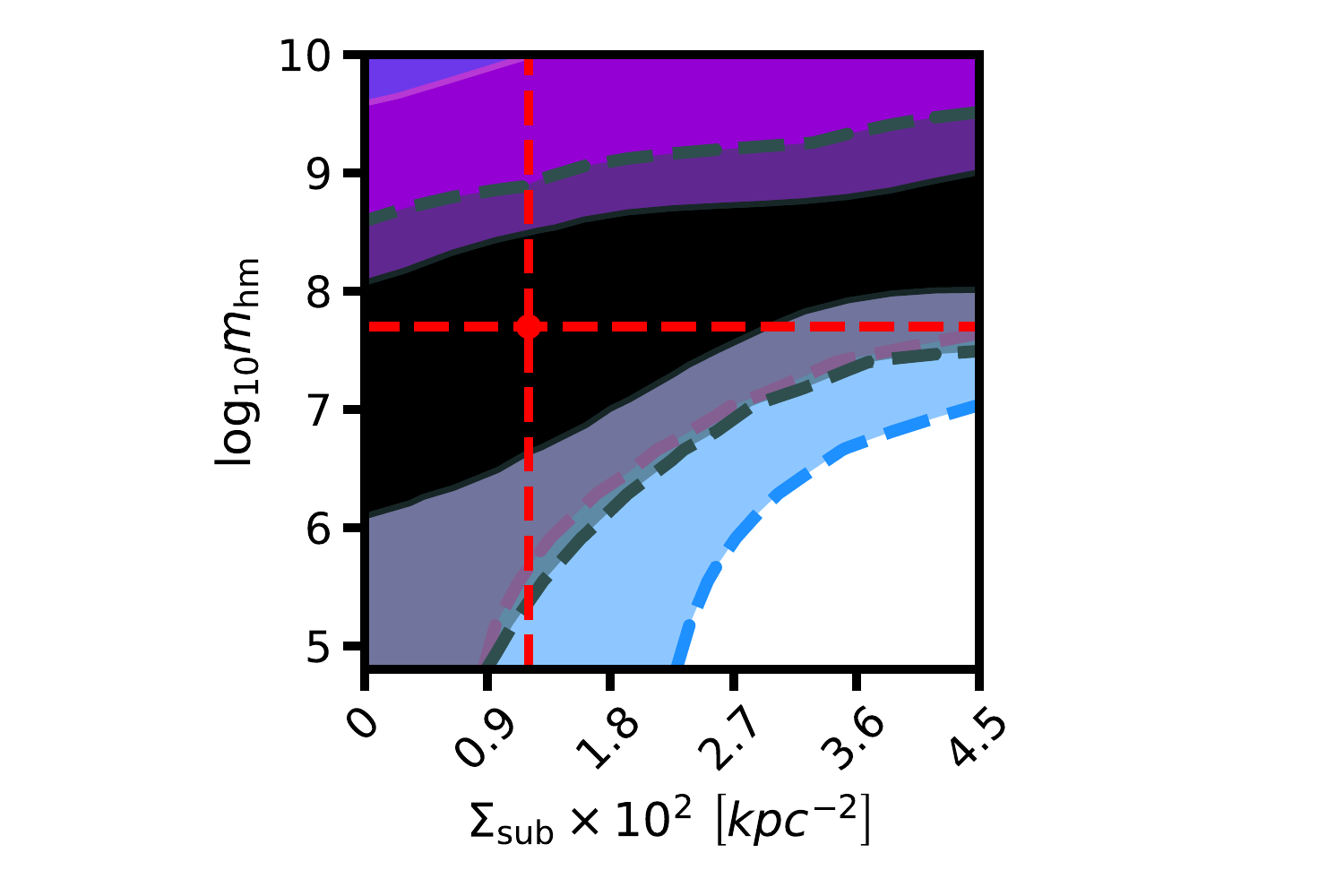}
		\includegraphics[clip,trim=2cm 0cm 3.5cm
		0.5cm,width=.48\textwidth,keepaspectratio]{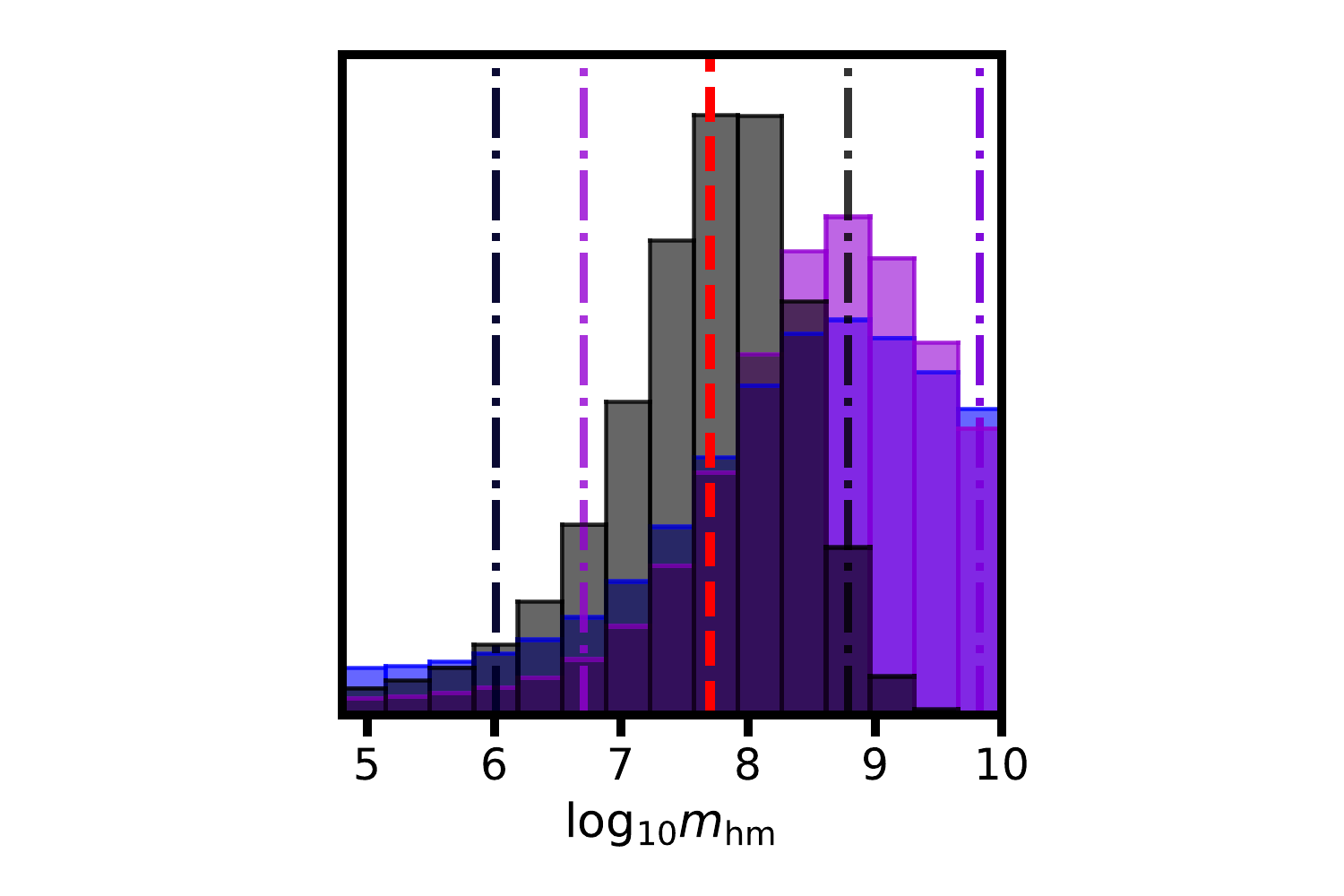}
		\caption{\label{fig:WDM_1} Inference on a WDM mass function with a half-mode mass of $10^{7.7} \msun$ ($m_{\rm{thermal}} = 5.4 \ \rm{keV}$), with the same color scheme as Figure \ref{fig:CDM_1}. As in Figure \ref{fig:CDM_1}, we marginalize over the parameters listed in \ref{tab:params} and over various degrees of flux uncertainty, and the color scheme is the same as in Figures \ref{fig:CDM_1} and \ref{fig:fluxuncertainties_CDM1}. For flux uncertainties of $2\%$, $4\%$, and $6\%$, we favor WDM with $\mhm > 10^{7.7} \msun$ over CDM with likelihood ratios of 22, 30, and 8, respectively.}
	\end{figure*}	
	\begin{figure*}
		\includegraphics[clip,trim=2cm 0cm 3.5cm
		0.25cm,width=.48\textwidth,keepaspectratio]{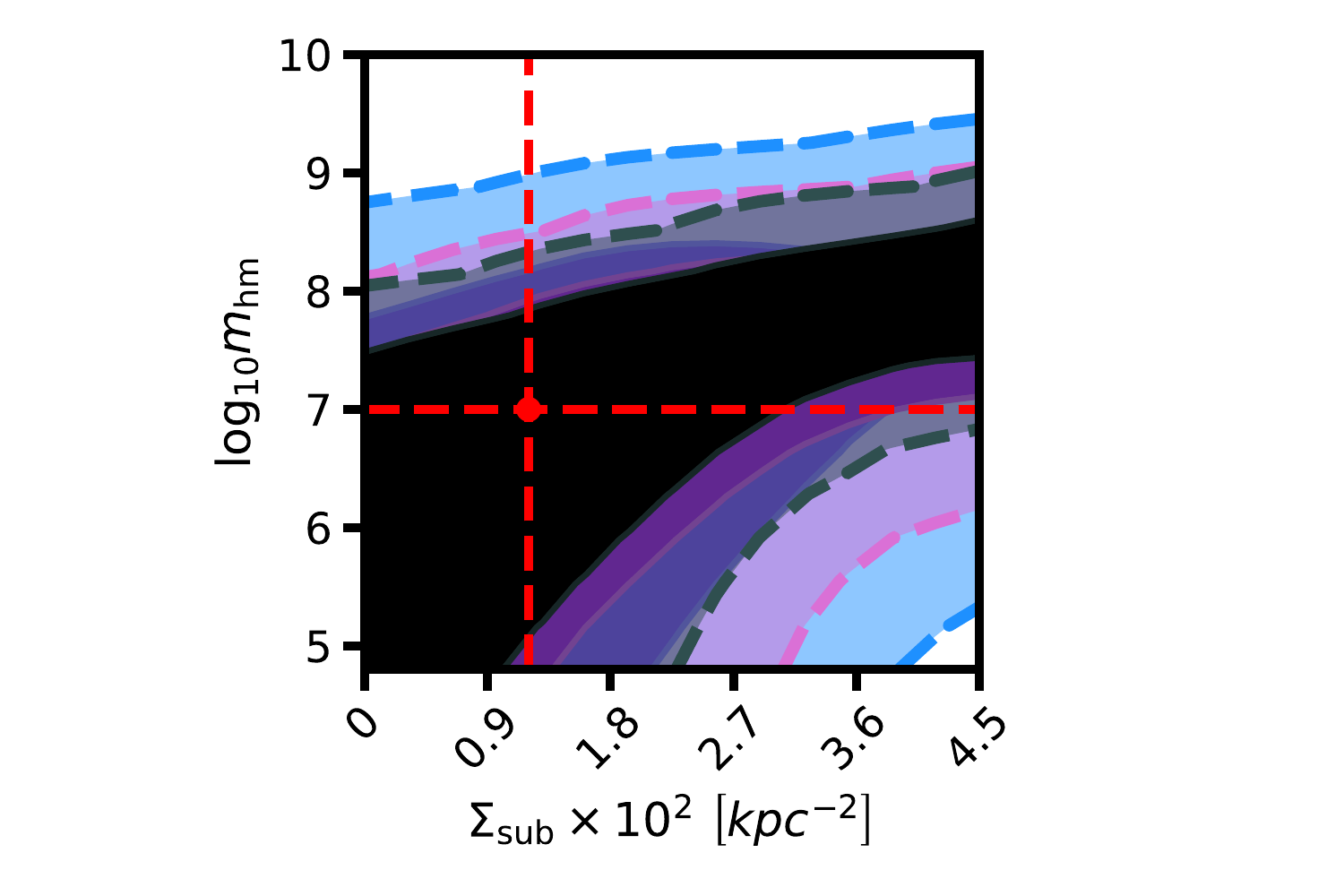}
		\includegraphics[clip,trim=2cm 0cm 3.5cm
		0.5cm,width=.48\textwidth,keepaspectratio]{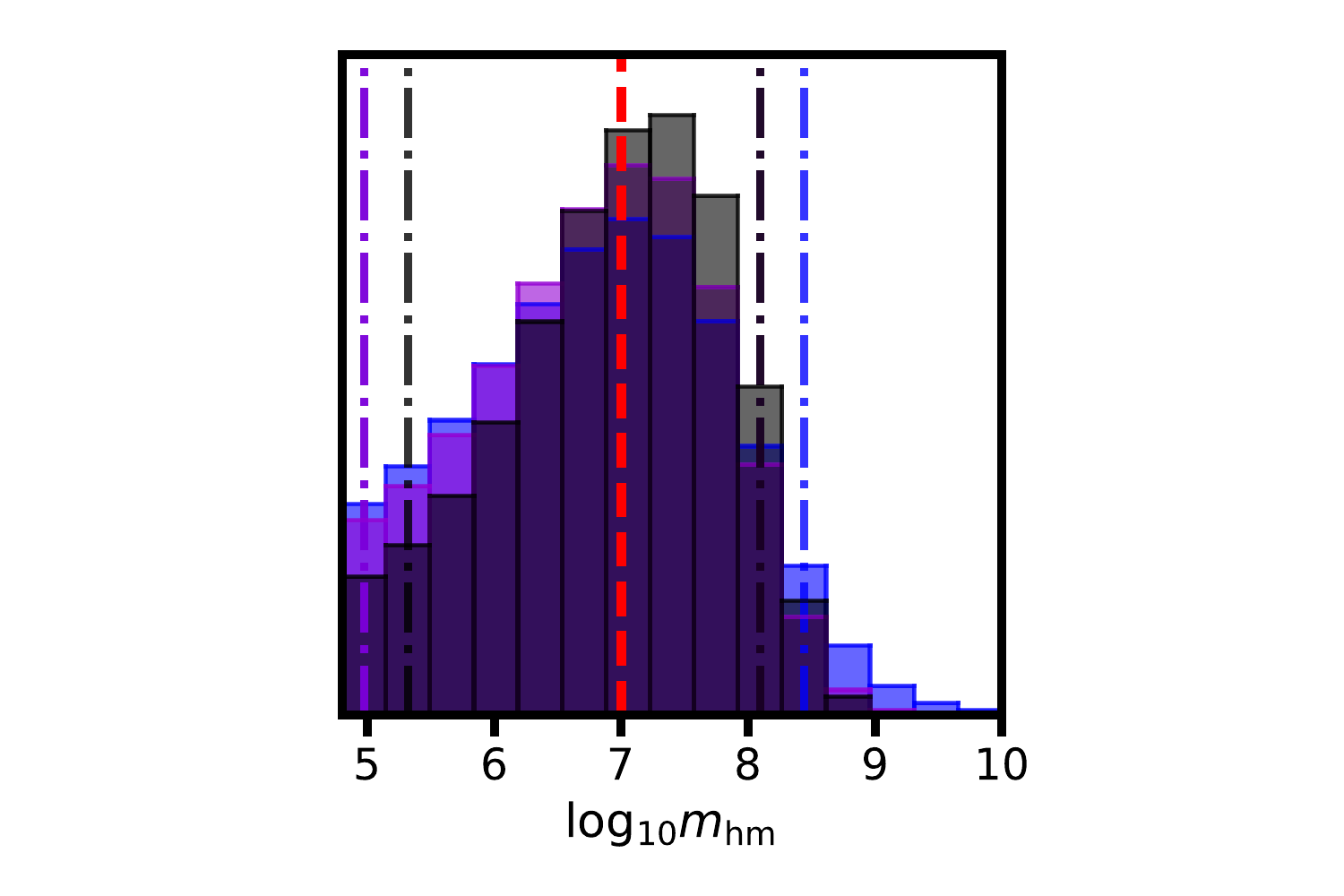}
		\caption{\label{fig:WDM_2} Inference on a WDM mass function with a half-mode mass of $10^{7} \msun$ ($m_{\rm{thermal}} = 8.2 \ \rm{keV}$)., marginalized over the parameters listed in Table \ref{tab:params}, with the same color scheme as Figure \ref{fig:CDM_1}. For each degree of uncertainty in image fluxes, the peak of the posterior coincides with the location of the the turnover at $10^{7}\msun$, but the width of the distributions increases. With uncertainties of $2\%$, $4\%$, and $6\%$ we favor WDM mass functions with $\mhm > 10^{7} \msun$ over CDM with likelihood ratios of 4:1, 3:1, and 2:1, respectively.}
	\end{figure*}	
	This section presents the results of our analysis, in which we test the forward modeling machinery described in the previous section to constrain dark matter properties. We discuss how measurement and modeling uncertainties affect the precision of constraints on both CDM and WDM mass functions, and make projections for the constraints on the half-mode mass. We explicitly consider 4 models: Two CDM cases with a different normalization of the subhalo mass function $\Sigma_{\rm{sub}}$, and two WDM cases with half-mode masses of $10^{7.7} \msun$, and $10^{7} \msun$. 
	
	\subsection{Joint inference on model parameters}
	\label{ssec:inference}
	\begin{figure}
		\includegraphics[clip,trim=0cm 0cm 0cm
		0cm,width=.46\textwidth,keepaspectratio]{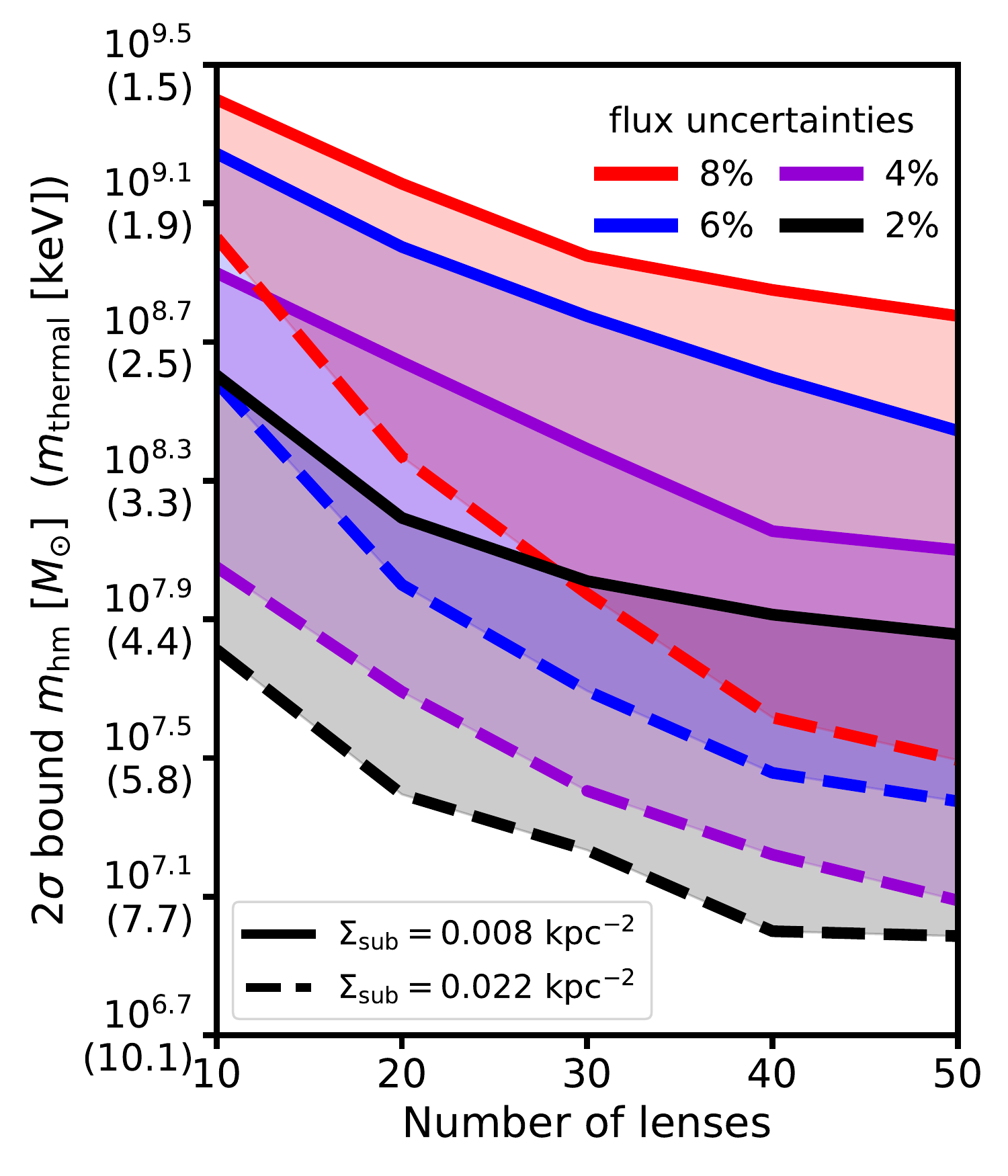}
		\caption{\label{fig:2sigmamhm} Forecasts for the constraints on the half-mode mass as a function of the number of lenses, including line of sight halos and subhalos of the main deflector. Black, purple, blue, and red colors denote flux uncertainties of $2\%$, $4\%$, $6\%$, and $8\%$. The solid line corresponds to a normalization $\Sigma_{\rm{sub}} = 0.008 \rm{kpc^{-2}}$, while dashed lines correspond to $\Sigma_{\rm{sub}} = 0.022 \rm{kpc^{-2}}$. The y-axis labels represent the $2 \sigma$ bound on $\mhm$, with the mass of the corresponding thermal relic dark matter particle in parentheses. Models with more subhalos (dashed lines), and hence more signal, are more resilient to flux uncertainties than models with fewer lens planes subhalos (solid lines) and produce stronger constraints on $\mhm$. }
	\end{figure}	
	Beginning with the CDM mass functions, in Figure \ref{fig:CDM_1} we show posterior distributions for all the parameters sampled in the forward model for a CDM mass function with a normalization of $\Sigma_{\rm{sub}} = 0.018 \ \rm{kpc^{-2}}$. As described in Section \ref{sec:simsetup}, we add flux perturbations of $2 \%$, $4 \%$, and $6 \%$ the mock data and model fluxes to simulate measurement errors, and additional sources of flux uncertainty that stem from lens modeling. We marginalize over ten realizations of these flux perturbations to reduce shot noise in the posterior distributions. 
	
	The boost in signal from the line of sight halos permits 2$\sigma$ bounds on the half-mode mass that range between $\mhm < 10^{7.1} \msun$, or a 7.9 keV thermal relic particle, to $\mhm < 10^{8.8} \msun$ (2.4 keV) as statistical measurement errors and modeling uncertainties in image fluxes increase from $2 \%$ to $6\%$. This rapid erosion of constraining power underscores the necessity of accurately measuring image fluxes, and accurate lens model predictions for these observables. 
	
	The most visibly striking covariance in Figure \ref{fig:CDM_1} exists between $\Sigma_{\rm{sub}}$ and $\mhm$ (see also Figure \ref{fig:fluxuncertainties_CDM1}). Physically, this feature corresponds to adding more substructure by increasingly the normalization, and subsequently removing some of the subhalos by raising the half-mode mass such that the total amount of flux perturbation remains relatively constant. Thus, above a sensitivity threshold of roughly $10^{6} \msun$, $\Sigma_{\rm{sub}}$ and $\mhm$ are positivity correlated. The opposite is true for $\Sigma_{\rm{sub}}$ and $\delta_{\rm{los}}$: the additional source of flux perturbation from extra line of sight structure is partially offset by reducing the number of lens plane subhalos, and these parameters are anti-correlated. Finally, there is weak evidence (notice the slightly tilted $2\sigma$ contours) for a positive correlation between the power-law slope of the macromodel $\gamma_{\rm{macro}}$ and the source size $\sigma_{\rm{src}}$. Without a priori knowledge of the true source size, the focusing power of a lens with a steeper mass profile makes larger background sources look smaller. Thus, a more extended background source is focused to the same size image by a steeper mass profile and these parameters are positively correlated. We emphasize that despite the covariance between parameters such as $\Sigma_{\rm{sub}}$ and $\mhm$, the data still constrains these parameters independently. The covariance affects the precision of the inference, but it does not result in completely unconstrained posterior distributions.  
	
	The normalization of the subhalo mass function $\Sigma_{\rm{sub}}$ plays an important role in the constraints on WDM and CDM models. Systems with more substructure are effectively weighted more than systems with fewer subhalos, and the strength of the constraints on $\mhm$ reflect this weighting. We illustrate this effect in Figure \ref{fig:fluxuncertainties_CDM1}, through comparison with Figure \ref{fig:CDM_1}. The former has $\Sigma_{\rm{sub}} = 0.01 \ \rm{kpc}^{-2}$, while the latter has nearly twice as many lens plane subhalos with $\Sigma_{\rm{sub}} = 0.018 \ \rm{kpc}^{-2}$. The constraints on $\mhm$ are weaker for the simulation with less substructure, because the data contains less signal. Due to the covariance between $\Sigma_{\rm{sub}}$ and $\mhm$, a significant portion of the volume of the posterior lies in high $\Sigma_{\rm{sub}}$, high $\mhm$ parameter space, which results in a peak in the marginalized constraint on $\mhm$. Stronger theoretical priors on $\Sigma_{\rm{sub}}$, which take into account the role of halo mass, redshift, and tidal stripping, may improve constraints on $\mhm$ by breaking this covariance.
	
	It is possible that by extending the range of the prior on $\Sigma_{\rm{sub}}$ to higher values, the covariance between $\mhm$ and $\Sigma_{\rm{sub}}$ would result in weaker constraints on the half-mode mass. However, extending the prior in this manner would imply a degree of ignorance surrounding the parameter $\Sigma_{\rm{sub}}$ that would likely be exaggerated given the current state of numerical simulations of dark matter halos and their substructure \citep{Benson12,Wheeler++18,Bozek++19,Lovell++18}. Keeping the width of the prior fixed, we implicitly assume that one may predict $\Sigma_{\rm{sub}}$ for each lens halo to within the width a factor of 4.5, or the width of the prior on $\Sigma_{\rm{sub}}$. 
	
	In Figures \ref{fig:WDM_1} and \ref{fig:WDM_2}, we show the constraints on WDM mass functions with $\mhm$ of $10^{7.7} \msun$ and $10^{7} \msun$, which correspond to thermal relic dark matter particles of 5.1 and 8.2 keV, respectively. Both datasets have $\Sigma_{\rm{sub}} = 0.012 \ \rm{kpc^{-2}}$. As in Figures \ref{fig:CDM_1} and \ref{fig:fluxuncertainties_CDM1}, we marginalize over every parameter listed in Table \ref{tab:params}, but focus only on the joint distribution of $\Sigma_{\rm{sub}}$ and $\mhm$. We see evidence for a turnover in the mass function, even though it lies below $10^8 \msun$. When interpreting the marginalized posteriors for $\mhm$ in cases where there is a clear peak in WDM territory, we use the relative likelihood between the lowest $\mhm$ bin (at $10^{4.8}\msun$) and the peak of the posterior as a summary statistic, since the statement regarding the $2 \sigma$ confidence interval depends on the width of the prior. \footnote{Sometimes, inference on CDM mass functions results in a posterior distribution peaked around some value of $\mhm$, due to the covariance between $\mhm$ and other parameters. This effect is visible in Figure \ref{fig:fluxuncertainties_CDM1}. In the case of Figure \ref{fig:fluxuncertainties_CDM1}, the maximum likelihood ratio between WDM and CDM with uncertainties of $2\%$ equals two.} 
	
	In the case of $\mhm=10^{7.7}\msun$, with flux uncertainties of $2\%$, $4\%$, and $6\%$, we favor WDM mass functions with $\mhm > 10^{7.7} \msun$ over CDM with relative likelihoods of 22:1, 30:1, and 8:1, respectively \footnote{The increase from 22 to 30 is likely due to shot noise.}. With uncertainties of $4\%$ and $6\%$, the posterior distributions of $\mhm$ shift towards higher masses, and the posteriors no longer resolve the position of the turnover in the mass function and mass-concentration relation. The shift to higher values of $\mhm$ is a consequence of the weak signal produced by very warm mass functions with a paucity of small-scale structure. Increased flux uncertainties wash out the information from the `weak signal' regime of parameter space with $\mhm > 10^{7.7} \msun$, and the constraints on this region of parameter space deteriorate because the data itself lies in this `weak signal' regime. This reasoning is similar to the interpretation of $\Sigma_{\rm{sub}}$ as an information scaling parameter for CDM mass function: like a CDM mass function with a high normalization, a `colder' WDM mass function produces more significant flux perturbation events, and is more resilient to increased uncertainties in image fluxes. If this reasoning is correct, we should expect the posteriors on $\mhm$ for `colder' WDM mass function to remain relatively stationary, modulo an increased variance, after adding perturbations to the image fluxes. 
	
	This effect is seen in Figure \ref{fig:WDM_2}, which has $\mhm=10^{7}\msun$. The shift of the posterior distributions towards higher masses as flux uncertainties increase does not happen in this case because the WDM mass function with $\mhm = 10^{7} \msun$ produces stronger perturbations in the data than the warmer, `weak signal' model with $\mhm = 10^{7.7}\msun$. This is because the halos are both more numerous and more concentrated that the WDM model with $m_{\rm{hm}} = 10^{7.7}\msun$. In turn, the stronger signal survives additional flux uncertainties, and is sufficient to constrain very warm mass functions. The locations of the peaks of the posteriors coincide with the true value of $\mhm$, but the width of the distributions widen. In this case, we favor WDM mass functions with $\mhm > 10^{7} \msun$ over CDM mass functions with relative likelihoods of 4:1, 3:1, and 2:1 with flux uncertainties of $2 \%$, $4 \%$, and $6 \%$, respectively. The fact that we statistically favor WDM models over CDM models suggests that we could infer a turnover in the mass function at $\mhm = 10^{7}\msun$ (or an 8.2 keV WDM particle) at higher significance with a larger sample of quads.
	
	\subsection{Marginalized constraints on the free-streaming length}
	\label{ssec:marginalized}
	The posterior distributions in Figures \ref{fig:CDM_1} and \ref{fig:fluxuncertainties_CDM1} give a sense for how the constraints on the half-mode mass in WDM models depends on the precision with which one measures image fluxes and predicts them with lens models, and on parameters such as the normalization of the subhalo mass function. To take into account sample variance, in Figure \ref{fig:2sigmamhm} we plot the marginalized constraints on the half-mode mass as a function of the number of lenses, $\Sigma_{\rm{sub}}$, and flux measurement uncertainties of $2\%$, $4\%$, $6\%$, and $8\%$. We plot the bounds on $\mhm$ for both a high $\left(\Sigma_{\rm{sub}} = 0.02 \rm{kpc^{-2}}\right)$ and low $\left(\Sigma_{\rm{sub}} = 0.008 \rm{kpc^{-2}}\right)$ normalization of the subhalo mass function. To produce these curves, we compute 200 bootstraps of 50 lenses, and average over many realizations of flux uncertainties. 
	
	With a sample of 50 lenses it will be possible to probe below $10^8 \msun$ in the halo mass function, to a degree that depends on the amount of substructure in the main deflector, measurement precision of image fluxes, and precise lens model predictions for this observable. With control over image fluxes at the level for $4\%$, routinely achieved at present \citep{Nierenberg++14,Nierenberg++17}, the bounds on $\mhm$ with 50 quads range between $10^{7.1} - 10^{8.1}\msun$ for values of $\Sigma_{\rm{sub}}$ of 0.01 and 0.022 $\rm{kpc}^{-2}$, respectively. With more precise predictions of $\Sigma_{\rm{sub}}$ made on a lens-by-lens basis, these bounds may improve.  We also note that future surveys, such as LSST, WFIRST, and Euclid, will discover hundreds of quads \citep{OguriMarshall10}, so the sample of available quads will eventually be much larger than 50. 
	
	\section{Summary and conclusions}
	\label{sec:conclusion}
	We have presented a method to perform Bayesian inference on the halo mass function through a forward modeling analysis of image flux ratios in quadruply imaged quasars. We model the contribution from line of sight halos, which boost the signal per lens and permit stronger constraints on the properties of dark matter with fewer systems. We demonstrate the method with a sample of 50 quads, comparable in number to the currently observed sample size, and project the constraints on the free streaming length of a WDM particle under different degrees of flux measurement and lens modeling uncertainties, while marginalizing over parameters describing the size of the background source, the lens macromodel, and the amplitude of the line of sight halo mass function. 
	
	Our key results can be summarized as follows:
	
	\begin{itemize}
		\item With a sample of 50 quads, we are able to constrain the free streaming length of dark matter on scales below $10^8 \msun$. Assuming CDM, with mean subhalo mass function normalizations $\Sigma_{\rm{sub}} = 0.022 \rm{kpc^{-2}} \ \left(0.008 \rm{kpc^{-2}}\right)$ we forecast bounds on the half-mode mass of $10^{7} \ \left(10^{7.9}\right) \msun$, $10^{7.1} \ \left(10^{8.1}\right) \msun$, $10^{7.4} \left(10^{8.4}\right) \msun$, $10^{7.5} \ \left(10^{8.8}\right) \msun$ for flux uncertainties of $2 \% ,\ 4\%, \ 6\%, \ \rm{and} \ 8\%$, respectively. These $\mhm$ limits translate to bounds on the mass of thermal relic particles of 8.2 (4.4), 7.7, (3.8), 6.2 (3.1), 5.8 (2.4) keV. 
		\item Line of sight halos contribute substantially to the signal in flux ratios, even dominating the signal in lens systems with higher lens and source redshifts. However, the normalization of the subhalo mass function still plays a key role in scaling the information content per lens, with higher values of this parameter translating into tighter constraints on the mass function. The half-mode mass is also covariant with the normalization, which affects the marginalized constraints on this parameter. These features underscore the importance of theoretical work to predict the projected surface mass density of substructure inside galactic halos with accurate models of baryonic feedback and tidal stripping.
		\item In the case that dark matter is warm, we are able to infer the location of the turnover in the mass function with 50 quads, even if it lies below $10^8 \msun$. With a half-mode mass of $10^{7.7} \msun$, which corresponds to a 5.1 keV thermal relic particle, we favor WDM mass functions with $\mhm > 10^{7.7}\msun$ over CDM with relative likelihoods of 22:1, 30:1 and 8:1 for flux uncertainties of $2 \% ,\ 4\%$, and $6\%$, respectively. With the same set of flux uncertainties and a half-mode mass of $10^{7} \msun$, we favor WDM with $\mhm > 10^{7}\msun$ over CDM with relative likelihoods of 4:1, 3:1, and 2:1. These constraints will likely improve with additional lenses, which suggests that a future large sample of quads could be used to infer a turnover in the halo mass function at $10^{7} \msun$ at high statistical significance.  
	\end{itemize}
	
	Our work is broadly consistent with other studies of the line of sight contribution in substructure lensing. For instance, by ray tracing through N-body simulations, \citep{Xu++12} compare the frequency of flux anomalies induced by line of sight versus main lens halos, and reach the conclusion that line of sight halos contribute at the same level as subhalos. More recently, \citep{Despali++18} analyze the role of line of sight halos in the context of gravitational imaging. This method differs somewhat from this analysis in that it aims to detect individual halos along the line of sight, and in the main lens plane, but the authors reach a similar conclusion: the line of sight contribution substantially boosts the signal per lens. In terms of relative numbers, line of sight halos can outnumber lens plane subhalos by a factor of 2-25, depending on the normalization of the subhalo mass function, and the lens and source redshifts. However, the most robust metric of the influence of line of sight halos comes from the resulting constraints on the half-mode mass. Differences in the treatment of the subhalo mass function, background source size, lens macromodel, and the lens redshift distribution complicate a simple comparison between this work and the results obtained in \citet{Gilman++18} by modeling only subhalos of the main deflector. With that said, the constraints obtained in this work by including line of sight halos, at the level of $10^{7} \msun$, are stronger by half an order of magnitude to one full order of magnitude over those obtained by \citet{Gilman++18}.
	
	The strength of the constraints on WDM models depend sensitively on the normalization of the subhalo mass function. This is partly due to the interpretation of the normalization as scaling the information content per lens, and also due to the covariance between the normalization and the half-mode mass, although we stress that despite this covariance both parameters can be constrained independently. This highlights the importance of refining theoretical predictions for the value of the normalization, accounting for halo mass, redshift, and the destruction of subhalos by tidal stripping. To this end, observables from each lens system, such as the central velocity dispersion, half-light radius, redshift, etc. may be used to inform the prior on the normalization and thus further improve the inferred posterior with actual data. 
	
	The macromodel used to describe the mass profile of the main deflector plays a key role in this analysis. Several studies demonstrate that simple parameterizations sometimes fail to fit the flux ratios of substructure-less mass profiles, leading to `artificial' flux ratio anomalies in the sense that they do not derive from dark matter substructure \citep{Gilman++17,Hsueh++18}. However, we note that these cases are dominated by the presence of undetected stellar disks, which are rare in the early-type galaxies that dominate the lensing cross section \citep{Auger++10,Shankar++17}. Also, we point out that identifying morphologically complexity in the main deflector and modeling it can remove these `artificial' anomalies \citep{Hsueh++16}. While we do not explicitly account morphologically complex deflectors in this work, we do allow some freedom in the macromodel by marginalizing over the power law slope, and account for additional variations in the image fluxes as high as $8\%$ that would result from marginalizing over additional macromodel parameters in the forward model. 
	
	Finally, we note that the formalism we present naturally accommodates other parameterizations of the halo mass function, and density profile for individual objects. 
	
	\section*{Acknowledgments}
	We thank Francis-Yan Cyr-Racine, Giulia Despali, Chuck Keeton, Stacy Kim, Alex Kusenko, Leonidas Moustakas, Annika Peter, and Simona Vegetti for helpful suggestions and interesting discussions throughout the course of this project. We also thank the anonymous referee for comments that improved the quality of this work. 
	
	DG, TT, and SB acknowledge support by the US National Science Foundation through grant AST-1714953. DG, TT, SB and AN acknowledge support from HST-GO-15177. Support for Program number GO-15177 was provided by NASA through a grant from the Space Telescope Science Institute, which is operated by the Association of Universities for Research in Astronomy, Incorporated, under NASA contract NAS5-26555. AN acknowledges support from the NASA Postdoctoral Program Fellowship. 
	
	This work used computational and storage services associated with the Hoffman2 Shared Cluster provided by the UCLA Institute for Digital Research and Education's Research Technology Group. This work also used computational and storage services associated with the Aurora and Halo super computers. These resources were provided by funding from the JPL Office of the Chief Information Officer. 
	
	\bibliographystyle{mnras}
	\bibliography{LOS_forward}
	
	\appendix
	
	\section{\bf Implementing the two-halo term}
	\label{app:A}
	The two-halo term describes an excess of matter (relative to the mean density of the universe) near a large halo, or a peak in the density field. It is evaluated using the software package \textsc{colossus} \citep{Diemer17}, and takes the form
	\begin{eqnarray}
		\xi_{\rm{2halo}} \left(r, M, z\right) = b\left(M,z\right) \xi_{\rm{lin}}\left(r, z\right)
	\end{eqnarray}
	where $b\left(M,z\right)$ is the halo bias around a mass $M$, computed with the model presented by \citet{Tinker++10}, and
	\begin{equation}
		\xi_{\rm{lin}}\left(r,z\right) = \frac{1}{2 \pi^2} \int_0^\infty k^2 P(k,z) \frac{\sin(kr)}{kr} dk
	\end{equation}
	is the linear matter-matter correlation function at a distance r. While in principle WDM free-streaming should affect the linear power spectrum $P\left(k,z\right)$, we do not model this effect.
	
	We define a boost parameter $\beta$ in terms of $\xi_{\rm{2halo}}$ as 
	\begin{equation}
		\label{eqn:boost}
		\beta \left(M,z\right) = \frac{2}{r_{\rm{max}} - r_{\rm{min}}} \int_{r_{\rm{min}}}^{r_{\rm{max}}} \xi_{\rm{2halo}} \left(r^{\prime}, M, z\right) dr^{\prime}
	\end{equation} 
	where M denotes the parent halo mass, and the factor of 2 accounts for symmetry around the parent halo. We choose $r_{\rm{min}} = 0.5 \rm{Mpc}$ and $r_{\rm{max}} = 10 \rm{Mpc}$, which captures most of the contribution from the correlation function while omitting the contribution from regions inside the virial radius of the parent halo. Defining $A_0\left(z\right)$ as the normalization of the halo mass function in the lens plane closest to the main lens halo, we incorporate the two halo term by taking $A_0\left(z\right) \rightarrow \left(1+\beta\right)  \times A_0\left(z\right)$, and add these halos at the main lens redshift. 
	
\begin{figure}
		\includegraphics[clip,trim=0cm 0.5cm 0cm
		0cm,width=.46\textwidth,keepaspectratio]{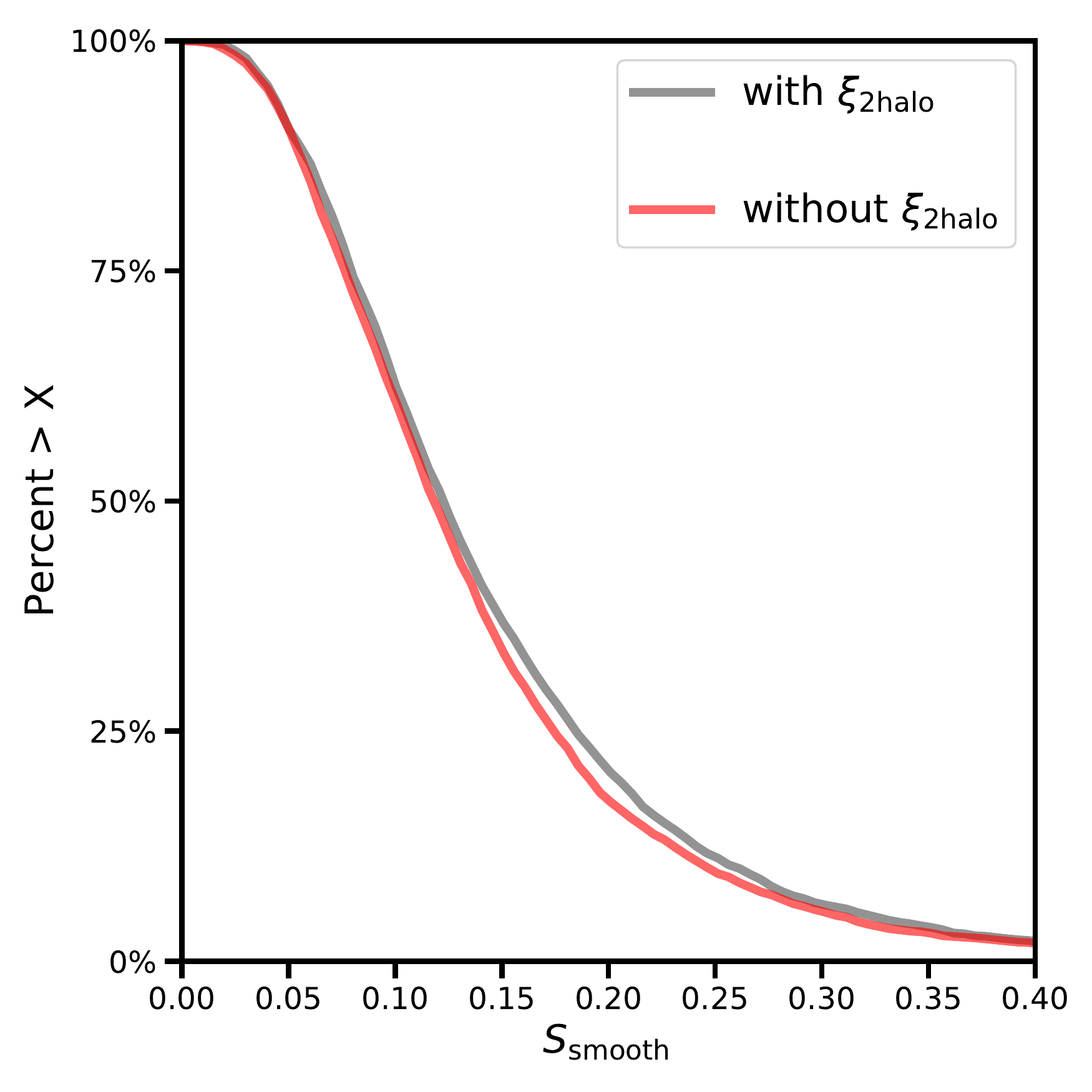}
		\caption{\label{fig:test2halo} Distributions of the summary statistic $S_{\rm{smooth}}$, which represents the amount of flux ratio anomaly with respect to a smooth lens model (see the discussion in Section \ref{sec:multiplanelensing}). The grey curve is computed with the two-halo contribution, and the red curve is computed without it. Both models include only line of sight halos to isolate the contribution from $\xi_{\rm{2halo}}$. The largest difference between the curves, an offset of $4\%$, lies at $S_{\rm{smooth}} \sim 0.2$. }
	\end{figure}

	In Figure \ref{fig:test2halo} we plot the distribution of summary statistics $S_{\rm{smooth}}$ for a CDM mass function that includes the boost from the two-halo term, and one that does not. In both cases, we set to $\Sigma_{\rm{sub}} = 0$ to isolate the impact of the two-halo term. The lens and source redshifts are set at 0.6 and 2, respectively. The largest differences between the curves occurs at $S_{\rm{smooth}}\sim 0.2$, and is equal to $4\%$. We conclude that the contribution from $\xi_{\rm{2halo}}$ is at most at the level of a few percent, although this may increase if a larger halo mass than $10^{13}\msun$ is used to evaluate Equation \ref{eqn:boost}. 
	
	\section{\bf The Born approximation in substructure lensing}
	\label{app:B}
	
\begin{figure}
		\includegraphics[clip,trim=0cm 0.5cm 0cm
		0cm,width=.46\textwidth,keepaspectratio]{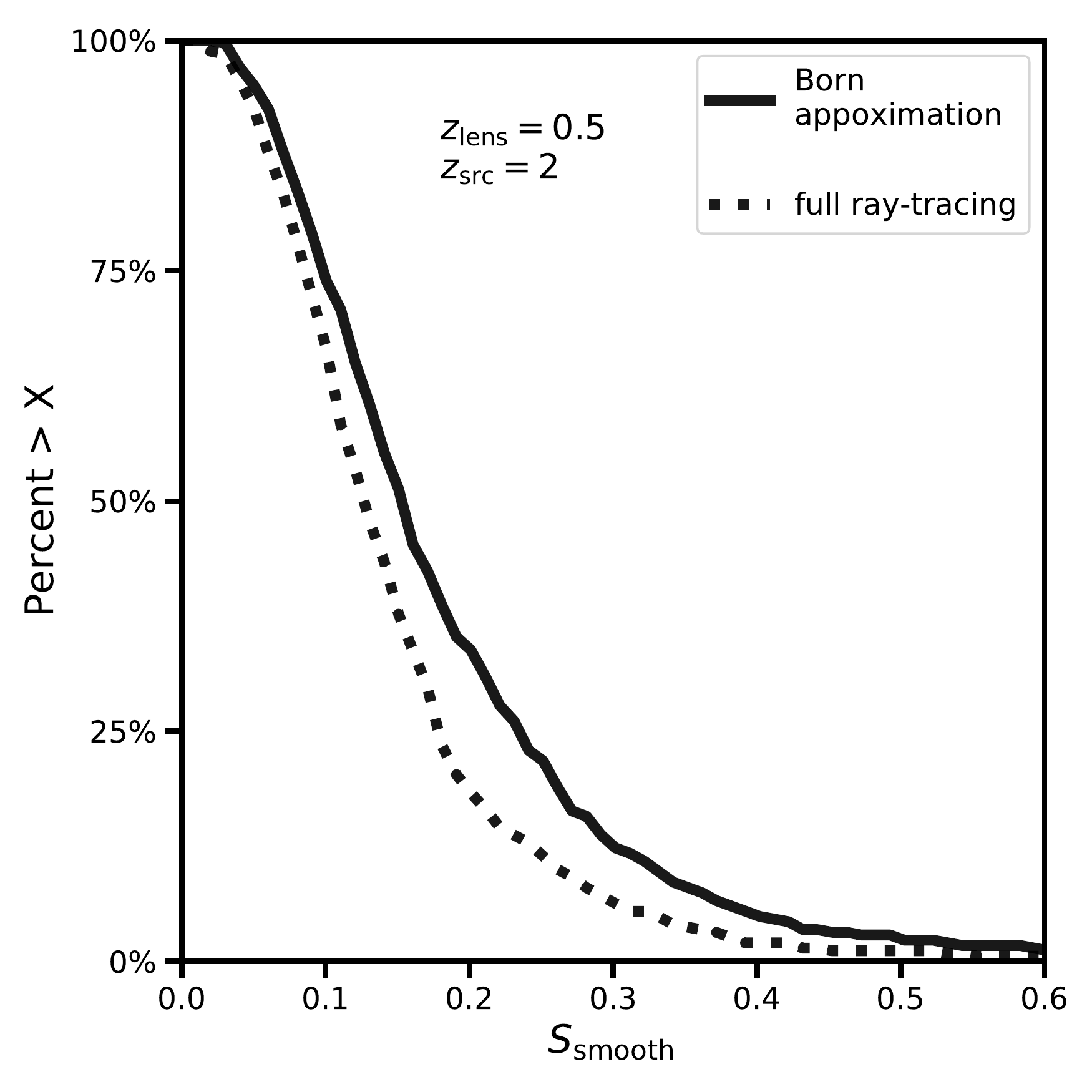}
		\caption{\label{fig:bornfluxes} The two curves show distributions of summary statistics computed with respect to a smooth lens model. The curves are computed for the same CDM mass function, with and without the use of the Born approximation. The disagreement between the two curves suggests that the Born approximation does not predict image flux ratios accurately enough to differentiate between dark matter models.}
	\end{figure}

	The Born approximation computes the deflection at each subsequent plane along an unperturbed path. This speeds up lensing computations since a full backwards ray-tracing routine is not required. In Figure \ref{fig:bornfluxes}, we compare the distribution of flux ratio anomalies computed with respect to a smooth lens model (see the discussion in Section \ref{ssec:lostats}) using the Born approximation, and through full multi-plane ray-tracing. The difference between the solid and dotted curves in the figure, which represent flux ratios computed with and without the Born approximation, respectively, is comparable to the difference of WDM and CDM mass functions in Figure \ref{fig:fluxdistributions}. Thus, we conclude that full multi-plane ray-tracing approach is required to accurately predict image flux ratios and probe dark matter on small scales. 
	
	\section{\bf A fast algorithm for multi-plane lensing computations}
	\label{app:C}
	\begin{figure}
		\includegraphics[clip,trim=1cm 4cm 1cm
		1cm,width=.48\textwidth,keepaspectratio]{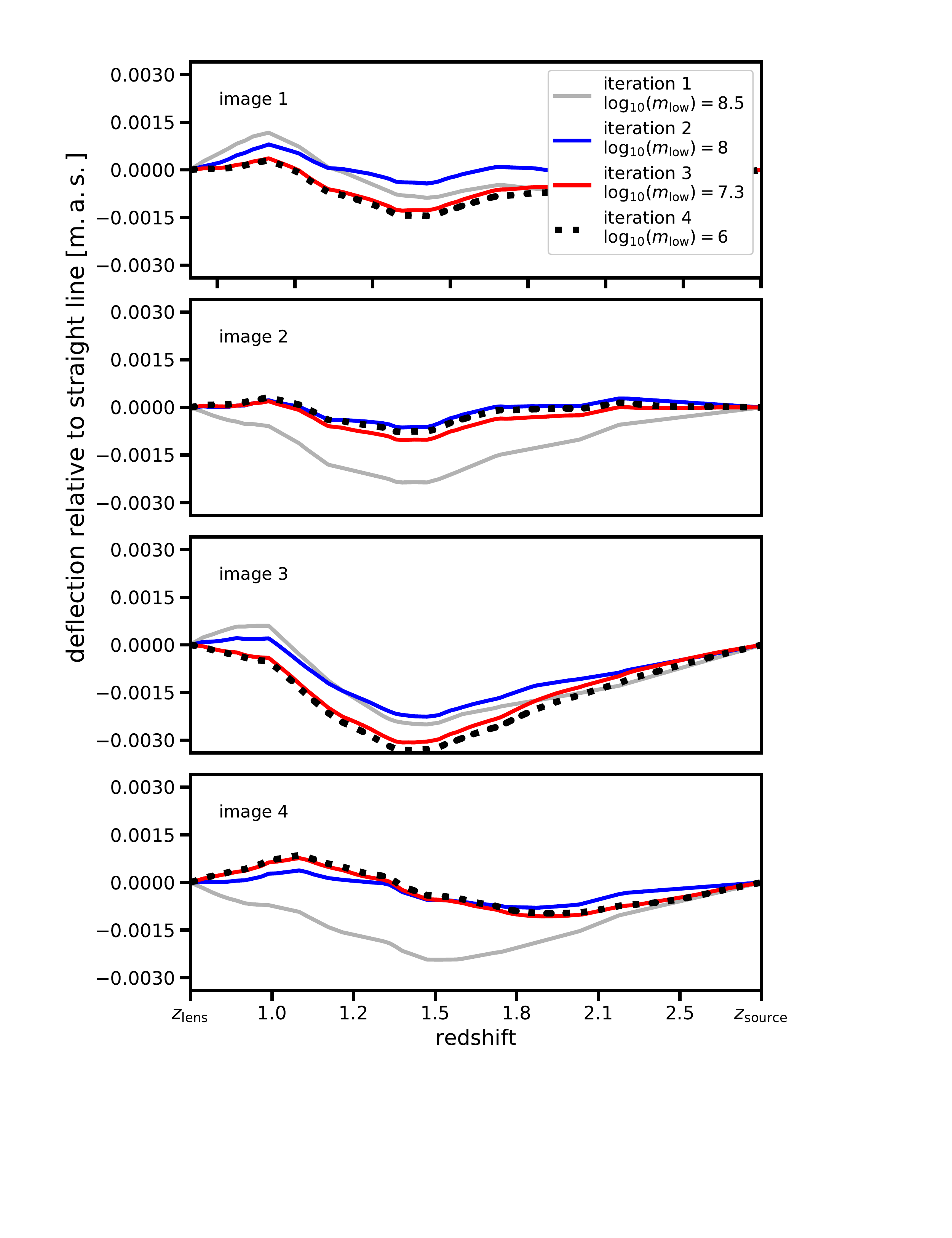}
		\caption{\label{fig:raytracealgorithm} A visualization of the perturbative ray tracing algorithm we use to optimize lens models with potentially thousands of line of sight halos. The panels show the path through the background field relative to a straight line for multiple iterations of the algorithm, in which progressively smaller halos are rendered in progressively smaller apertures around the path of the light rays. This procedure speeds up optimizations of lens models with line of sight halos by at least an order of magnitude.}
	\end{figure}	
	
	For each observed lens, our forward modeling approach requires finding a set of macromodel parameters that cast the four light rays in a quadrupole image system to the same location in the source plane. For a single realization, this typically requires hundreds to thousands of backwards ray-tracing computations. 
	
	This task is computationally light for models with halos only in front of and at the same redshift as the main deflector because the path through the foreground field of halos is not coupled to the deflections produced in the main lens plane (owing to the recursive nature of Equation \ref{eqn:raytracing}). Put differently, as soon as one specifies image positions on the sky and draws a realization of dark matter halos, the path through the foreground field is fully determined. In contrast, the path through the field of background halos is coupled to the deflections produced by the macromodel. The path through the background field therefore changes for each new proposal of macromodel parameters. This necessitates repeated computations of the potentially thousands of deflection angles of halos behind the main lens plane, which requires hundreds to thousands as many function evaluations as those needed in single plane lensing computations.
	
	We address this computational challenge by implementing a perturbative approach to lens model optimizations. First, we optimize the macromodel to fit image positions with only foreground halos and main deflector subhalos present. We denote this optimized lens model $\vec{m*}$. This proceeds quickly, since the macromodel deflection angles are not coupled to those from foreground and main lens plane halos. Next, we add the largest background halos with $m>10^8 \msun$, and {\textit{re-optimize}} $\vec{m*}$. Even though the deflections from these massive halos are coupled to those of the macromodel and need to be continuously re-evaluated during the optimization, since there are relatively few of them this proceeds fairly quickly. Next, we add halos in the range $10^{7.5} - 10^8 \msun$, but only in 300 m.a.s. apertures around the path of the rays computed with respect to $\vec{m*}$. Since the area in which we render these smaller halos is relatively small, and since the macromodel solution $\vec{m*}$ is already close to the true solution, this optimization also proceeds quickly. We iterate this process for progressively smaller halos until we reach $10^6 \msun$. 
	
	A visual representation of this process is presented in Figure \ref{fig:raytracealgorithm}, where we plot the path through the background halos relative to a straight line for subsequent iterations of the perturbative approach. After adding the $10^8 \msun$ background halos, the path through the background lens planes changes only slightly, which reflects the fact that these massive objects dominate the deflection field. 
	
	This procedure accomplishes the optimization of a macromodel with background halos 10-50 times faster than a naive optimization with all background halos included simultaneously. We test that the flux ratio statistics are identical to those obtained by ray tracing through full realizations without the perturbative approach implemented. We note that this algorithm is reminiscent of the Born approximation in that it initially neglects the presence of small deflections from subhalos along the line of sight, but differs fundamentally from the Born approximation in that the full non-linear coupling between every subhalo is eventually accounted for. 
	
	\section{\bf Convergence of posterior distributions}
	\label{app:D}
	
	We approximate the true posterior distributions for model parameters by retaining the top 1,500 samples (ranked by their summary statistics) out of the 600,000 realizations computed per lens. To test whether this procedure yields an accurate approximation to the true posterior distribution, we appeal to a certain feature of Approximate Bayesian Computing algorithms, namely, that the approximation to the true posterior distribution converges as the number of samples increases. We can therefore test for convergence by applying the same cut on the top 1,500 samples to an `under-sampled' model with only 400,000 realizations per lens, and check that the posterior distribution stays approximately fixed in place. We generate the sample of 400,000 by drawing the realizations randomly from the computed set of 600,000.
	
\begin{figure*}
		\includegraphics[clip,trim=2cm 0cm 1cm
		0cm,width=.9\textwidth,keepaspectratio]{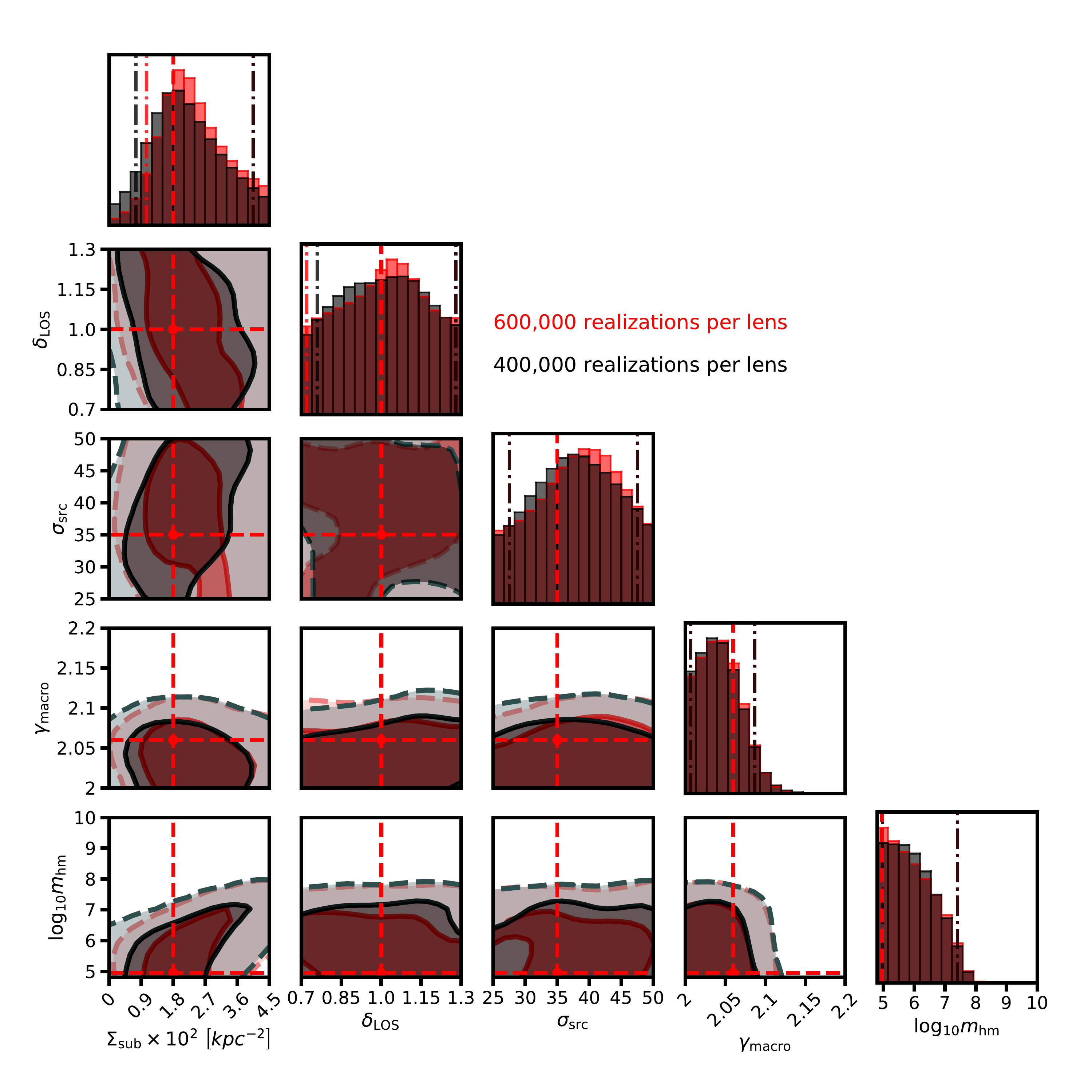}
		\caption{\label{fig:convergence_test} A convergence test for the forward model simulations. The overall agreement between the black and red distributions indicates that the posteriors we derive, and the numerical operations involved to produce them including the kernel density estimation, are robust to changes in the number of forward model samples per lens.}
	\end{figure*}

	We perform this test and plot the results in Figure \ref{fig:convergence_test}. While there is some movement in the $1 \sigma$ contours, the $2 \sigma$ contours trace each other closely. Importantly the constraints on the half-mode mass are the same between the two inferences, which is the most important criterion for our purpose of forecasting bounds on dark matter warmth. Finally, we note that ABC routines tend to yield conservative approximations to the true posterior distributions, in the sense that with more samples the volume of the resulting posterior distribution shrinks. This explains why black contours (400,000 samples) tend to cover more area than the red contours (600,000 samples). As additional forward model samples improve the precision of the inference, the constraints we present would only improve by computing additional realizations.  
	
\end{document}